\documentclass[a4paper,fleqn]{cas-dc}

\usepackage[authoryear]{natbib}

\usepackage{siunitx}

\def\tsc#1{\csdef{#1}{\textsc{\lowercase{#1}}\xspace}}
\tsc{WGM}
\tsc{QE}

\usepackage{tablefootnote}
\let\oldAA\AA
\renewcommand{\AA}{\text{\normalfont\oldAA}}

\begin{document}
\let\WriteBookmarks\relax
\def\floatpagepagefraction{1}
\def\textpagefraction{.001}

\shorttitle{Quantum Rate Electrodynamics in $\pi$-Conjugated Heterocyclic Molecules}    

\shortauthors{Paulo R. Bueno}  

\title [mode = title]{Quantum Rate Electrodynamics and Resonant Junction Electronics of Heterocyclic Molecules}

\author[1]{Edgar Fabian Pinzón Nieto}
\author[1]{Laís Cristine Lopes}
\author[1]{Adriano dos Santos}
\author[2]{Maria Manuela Marques Raposo}
\author[1]{Paulo Roberto Bueno}
\ead{paulo-roberto.bueno@unesp.br}
                      
\cormark[1]

\ead{paulo-roberto.bueno@unesp.br}

\affiliation[1]{organization={Departmnet of Engineering, Physics and Mathematics, Institute of Chemistry, São Paulo State University}, 
            city={Araraquara},
            postcode={14800-060}, 
            state={São Paulo},
            country={Brazil}}

\affiliation[2]{organization={Centre of Chemistry, University of Minho, Campus de Gualtar}, 
            city={Braga},
            postcode={4710-057}, 
            state={Braga},
            country={Portugal}}


\begin{abstract}[summary]
The quantum rate theory provides a framework to understand electron-transfer reactions by correlating the electron-transfer rate constant ($\nu$) with the quantum capacitance ($C_q$) and the molecular conductance ($G$). This theory, which is rooted in relativistic quantum electrodynamics, predicts a fundamental frequency $\nu = E/h$ for electron-transfer reactions, where $E$ is the energy associated with the density of states $C_q/e^2$. This work demonstrates the applicability of the quantum rate theory to the intermolecular charge transfer of push-pull heterocyclic compounds assembled over conducting electrodes. Remarkably, the observed differences between molecular junction electronics formed by push-pull molecules and the electrodynamics of electrochemical reactions on redox-active modified electrodes can be attributed solely to the adiabatic setting of the quantum conductance in push-pull molecular junctions. The electrolyte field-effect screening environment plays a crucial role in modulating the resonant quantum conductance dynamics of the molecule-bridge-electrode structure. In this context, the intermolecular electrodynamics within the frontier molecular orbital of push-pull heterocyclic molecules adhere to relativistic quantum mechanics, consistent with the predictions of the quantum rate theory.

\end{abstract}




\maketitle

\section{Introduction}\label{sec:introduction}

\begin{figure}
\centering
\includegraphics[width=7cm]{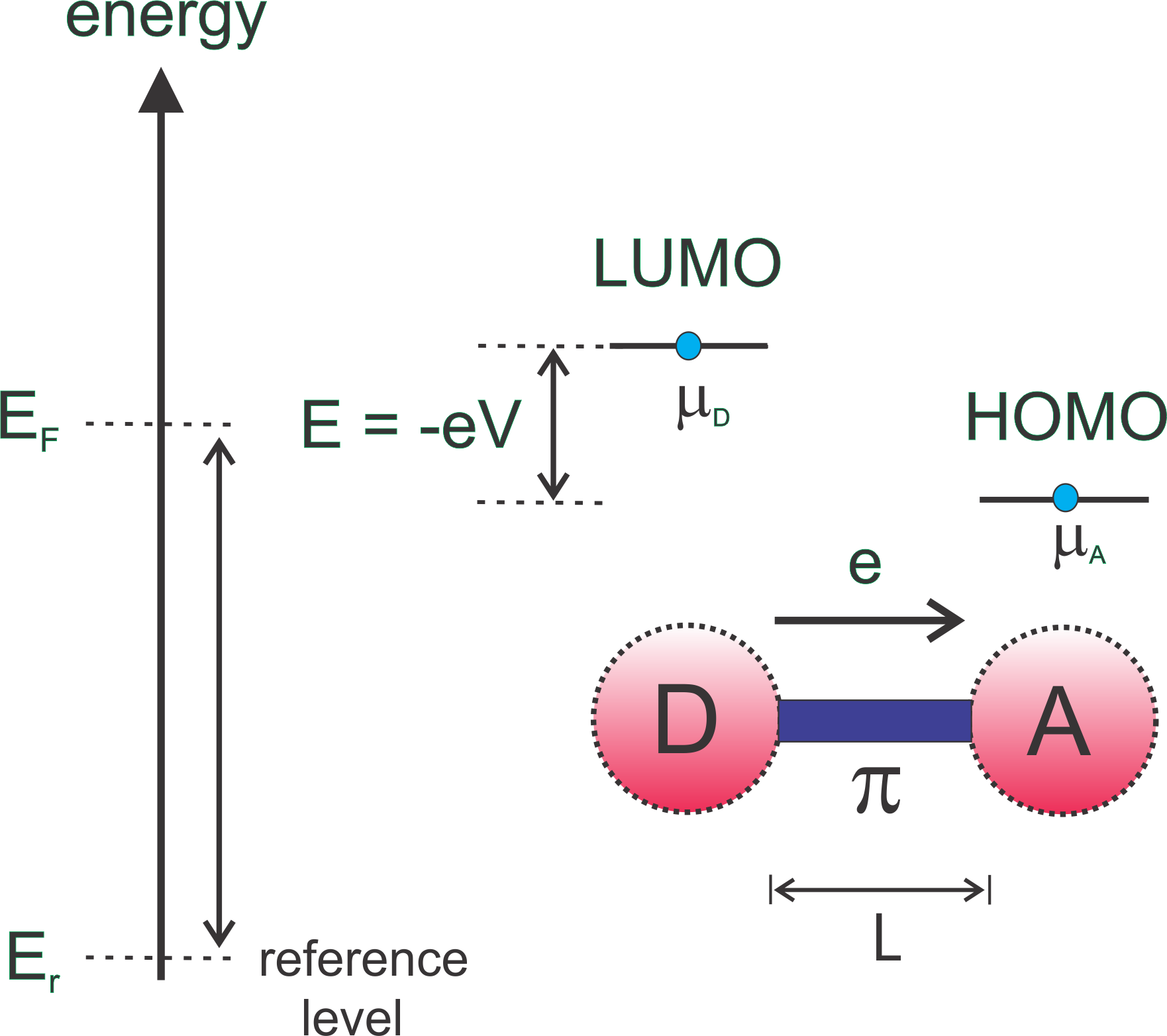}
\caption{The schematic representation (for the sake of simplicity without the consideration of the thermal broadening effect) of a $D-\pi-A$ intra-molecular charge resonant structure within the quantum rate theory, where $\mu_D$ and $\mu_A$ states for the chemical potential of $D$ (LUMO) and $A$ (HOMO) energy levels, respectively. Note that there is no energy barrier activation for the electron dynamics with a quantum channel of length $L$ within this molecular electronic structure, corresponding to an adiabatic setting for quantum electrodynamics within the quantum rate theory. The energy difference between $\mu_D - \mu_A$ is equivalent to the HOMO-LUMO energy gap and numerically it equates to $E = -eV = \mu - E_F$ that, within the quantum rate theory, corresponds to $h\nu = g_e e^2/C_q$, where $\nu = g_e G_0/C_q$ with $G_0 = g_s e^2/h$, where $V$ corresponds to the bias potential of the electrode regarding to $E_F/e$. $E_F$ states for the Fermi level of the resonant and adiabatically coupled $D$ and $A$ states for a $E_r$ reference level of energy, whereas $\mu$ is the chemical potential of the electrons in the electrode side of the junction. Note that the scheme of this figure considers $E = -eV = h\nu$ as an external excitation of electrons from HOMO to LUMO states that hence further relax to the HOMO level in a $D-\pi-A$ molecular structure.}
\label{fig:D-pi-A-scheme}
\end{figure}

Organic and molecular electronics represent modern fields of electronics in which the semiconducting properties of organic molecules are utilized. One notable application is the design of modern computer and mobile phone displays, exemplified by organic light-emitting diodes (OLEDs) \cite{Ma-2010}. The advantage of organic electronics over traditional ones, based on inorganic semiconducting materials, lies in their lower fabrication cost and formability \cite{Ma-2010}. Organic molecules containing delocalized $\pi$-electron systems have garnered widespread interest due to their applications as active materials in OLEDs, organic field-effect transistors, and photovoltaic devices \cite{Burevs-2014}. Within deslocalized $\pi$-electron compounds, the $\pi$-conjugated push-pull heterocyclic subclass of molecules is particularly important. These molecules contain electron donor ($D$) and acceptor ($A$) electronic groups bridged by a $\pi$-system, establishing a proper $D-A$ electron coupling and forming $D-\pi-A$ type resonant electronic structures~\cite{Fernandes-2018} (see Figure\ref{fig:D-pi-A-scheme}).

Synthetic chemistry and computational simulation studies have shown that replacing the benzene ring of a chromophore bridge with easily delocalized five-membered heteroaromatic rings, such as thiophene, pyrrole, and (benzo) thiazole, enhances the optoelectronic properties of push-pull molecules. Thiophene and pyrrole derivatives, as donors, substituted with appropriate acceptor groups, are promising candidates among such push-pull systems~\citep{Mayer-2012,Burevs-2014,raposo2006synthesis}. In recent years, researchers have investigated several push-pull heterocyclic molecules, including oligothiophene and thienylpyrrole derivatives. These compounds exhibit advantageous properties used in the manufacturing of semiconductor materials, sensitizers, compounds for optical data storage, photoswitches, chemosensory probes, and bioimaging probes~\citep{Mayer-2012,Burevs-2014,raposo2006synthesis,oliva2006structure, marin2014imidazoanthraquinone,castro2016synthesis,garcia2016fastest,Fernandes-2017,
gonccalves2022bioimaging,garcia2023photochromic}.

The utilization of $D-\pi-A$ heterocyclic structures is a viable strategy for designing high-performance organic semiconductors, exploiting the benefits of $D-A$ resonant intermolecular electron coupling provided by $\pi$-conjugated molecular bridges~\cite{Cai-2013}. This electronic coupling phenomenon, referred to in the literature as intramolecular charge-transfer (ICT), involves a relatively lower HOMO-LUMO energy gap ($\Delta E_{H-L}$), excited by photons in the visible region of the electromagnetic spectrum. Hence, the photo-absorption of push-pull heterocyclic compounds is tunable by the judicious choice of $D$ and $A$ groups and $\pi$-system constituent moieties of these compounds~\cite{Fernandes-2018,Fernandes-2021}. The design of $D-\pi-A$ push-pull molecules facilitates the development of optoelectronic devices. Electrochemical and optical methods are commonly employed to investigate the $\Delta E_{H-L}$ gap and electronic transitions between HOMO and LUMO states. Cyclic voltammetry is frequently used to estimate the energy accessibility of an electrode to the HOMO and LUMO of $D-\pi-A$ molecules in an electrolyte environment. Complementary absorption spectroscopy provides information about the ICT dynamics of $D-\pi-A$ compounds~\cite{Cardona-2011,Fernandes-2017}.

Electrochemical techniques have been indirectly used to access the electronic properties of organic heterocyclic compounds in contact with electrodes employed as a probe to observe the electronic communication. For example, a series of push-pull molecules consisting of bithiophene and thienothiophene moieties as $\pi$-bridges structures were characterized using cyclic voltammetry~\cite{Fernandes-2017, Fernandes-2018}.

In these types of experiments, the molecules are not chemically attached to the electrode but are free in a solvent environment, where electronic communication with the electrode is governed by the diffusion of the molecules to the electrode. In this type of experimental setting, the purpose is to estimate the HOMO and LUMO energy levels of the structures through the determination of the oxidation and reduction potentials of the compounds. The estimation of HOMO and LUMO energy levels is relevant for applications in nonlinear optics and dye-sensitized solar cells (DSSCs) owing to the achievement of the alignment between HOMO and LUMO levels of these organic molecules to semiconductor band gap levels is important for achieving better DSSC devices.

Furthermore, organic structures comprising the self-assembling of thiol-terminated dendritic oligothiophenes of varying lengths on gold surfaces~\cite{jiang2010monolayer} can effectively be used for blocking electron-transfer reactions occurring between the gold electrode and redox species added to the electrolyte. The employment of this type of molecular junction conducts to what is named blocking layers whose goal is to switch off the redox processes.

An additional example of the use of molecular junctions is related to the study of electron transport in organic electronic junctions, for instance, those comprising azobenzene molecules bridging carbon/gold electrodes. This type of junctions were investigated in a solid-state environment using impedance measurements~\cite{santos2020introducing} of the junctions. Recorded impedance spectra of the junctions permitted to study of the conductance and the electron injection rate responses also in adherence to quantum mechanical principles described by Eq.~\ref{eq:nu}, demonstrating that electronic and electrochemical principles are common at the molecular scale~\cite{bueno2018common} and follows similar quantum mechanical dynamics based on resistive-capacitive components of the junctions within in a molecular energy and length scale.

Although the above-discussed experimental methods and junction fabrication approaches provide insights into the electronic structure of organic compounds, the electronic properties of molecular junctions constructed using physical or chemical covalent attachment of push-pull molecules to metallic or semiconducting electrodes are rarely studied. In this work, the electronic properties of molecular junctions fabricated using push-pull heterocyclic compounds are characterized using time-dependent methods within a quantum rate theoretical viewpoint~\cite{Bueno-book-2018, Bueno-2023-3, Bueno-2020, Bueno-2023-1, Bueno-2023-2}. As a molecular junction model, we studied the 4\textit{H}-thieno[3,2-\textit{b}]pyrrole-5-carboxylic acid (\textbf{1,TPC}) molecule attached to a gold metallic electrode surface. The characterization of this \textbf{TPC}-Au junction was based on a spectroscopic low-energy perturbation of this molecular junction (quantum rate spectroscopy) in an electrolyte medium, providing an accurate characterization of this junction and revealing its low-energy quantum relativistic electrodynamics.

In the next section, we employ concepts of the quantum rate theory and a spectroscopic method based on the theory to characterize the \textbf{TPC}-Au junctions.

\subsection{Quantum Rate Theory}\label{sec:QRT}

The quantum rate theory, as elucidated by previous works~\citep{Bueno-book-2018, Bueno-2020}, correlates the electron transfer rate concept ($\nu$) with the molecular conductance ($G$), also known as quantum conductance, through the quantum capacitance ($C_q$). This theory provides a comprehensive framework for understanding the rate of electron transfer in various contexts, including electrochemical reactions~\citep{Sanchez-2022-1, Alarcon-2021}, the electrodynamics of graphene in electrolyte environments~\citep{Bueno-2022}, the respiration in Geobacter~\citep{Bueno-2015, Bueno-2024} and the intramolecular electrodynamics in $\pi$-conjugated heterocyclic molecules, as will be demonstrated here.

The quantum rate theory allows for modeling electron exchange between $D$ and $A$ states, elucidating the dynamics of intramolecular electrodynamics in push-pull molecules, akin to redox reactions~\citep{Bueno-2020,Bueno-2023-3}. In both scenarios, as verified in this study, the phenomenon adheres to quantum mechanical principles that cannot be accurately described using Schrödinger's non-relativistic quantum-wave methods but can be adequately addressed using quantum electrodynamics. The theory establishes a fundamental quantum rate ($\nu$) principle, expressed by the ratio between the reciprocal of the von Klitzing constant ($R_k = h/e^2$) and the quantum capacitance ($C_q$), as follows

\begin{equation}
\label{eq:nu}
\nu = \frac{e^2}{hC_q},
\end{equation}

\noindent which leads directly to the Planck-Einstein relationship

\begin{equation}
\label{eq:Planck-Einstein}
E = h\nu = \hbar \textbf{c}_* \cdot \textbf{k},
\end{equation}

\noindent where $\textbf{c}_*$ represents the Fermi velocity and $\hbar$ is the reduced Planck constant ($h/2\pi$). The relativistic dynamics predicted by this relationship conform to Dirac's relativistic wave mechanics, rather than the non-relativistic Schrödinger equation. This is evidenced by employing the de Broglie wave-particle duality relationship, in which the momentum is defined as $\textbf{p} = \hbar \textbf{k}$, being directly proportional to the wave-vector $\textbf{k}$ and leading to $E = \textbf{p} \cdot \textbf{c}_*$ in Eq.~\ref{eq:Planck-Einstein}. This constitutes an energy-momentum relationship that is described by relativistic quantum electrodynamics, which is the first success theory where quantum mechanics agree with relativistic principles~\cite{gluck2009classical,greiner2000relativistic, land2022relativistic}.

The quantum rate theory operates within the framework of quantum electrodynamics, reflecting the intrinsic relativistic character of electrons, but it does not mean that electrons are traveling with the speed of light as naively interpreted by those not familiar with these types of electrodynamics. The meaning is owing that electrons follow Eq.~\ref{eq:Planck-Einstein}, exhibiting a massless character (in which electrons are referred to as Fermionic particles) and thus behaving as waves within a constant Fermionic velocity $\textbf{c}_*$. This is evidenced whenever the electron's rest mass $m_0$ is null in the relativistic equation $E^2 = \left( \textbf{p} \cdot \textbf{c}_*\right)^2 + \left( m_0 c_*^2 \right)^2$, thus predicting the particular $E = \textbf{p} \cdot \textbf{c}_*$ quantum electrodynamics situation that conforms with Dirac's equation of the electrodynamics~\cite{Dirac-1928}. It is not novel that the electrodynamics of bidimensional single-layer graphene obeys this relativistic quantum electrodynamics~\cite{Novoselov2005a}, but there is experimental evidence~\citep{alarcon2021perspective, Bueno-2023-3, sanchez2022quantum, lopes2024electrochemical} supporting that the electrodynamics of diffusionless electrochemical reactions follows this massless Fermionic dynamics within a degenerate energy state $E = \textbf{p} \cdot \textbf{c}_* \propto e^2/C_q$. Specifically, in electrolyte environment, $E$ is proportional to $e^2/C_q$ by the $g_s$ and $g_e$ degeneracies, such as $E = g_s g_e (e^2/hC_q)$, where $g_s = 2$ represents electron spin degeneracy and $g_e = 2$ signifies the degeneracy associated with the electric-field screening effect of the electrolyte on molecular states.

\begin{figure}
\centering
\includegraphics[width=7cm]{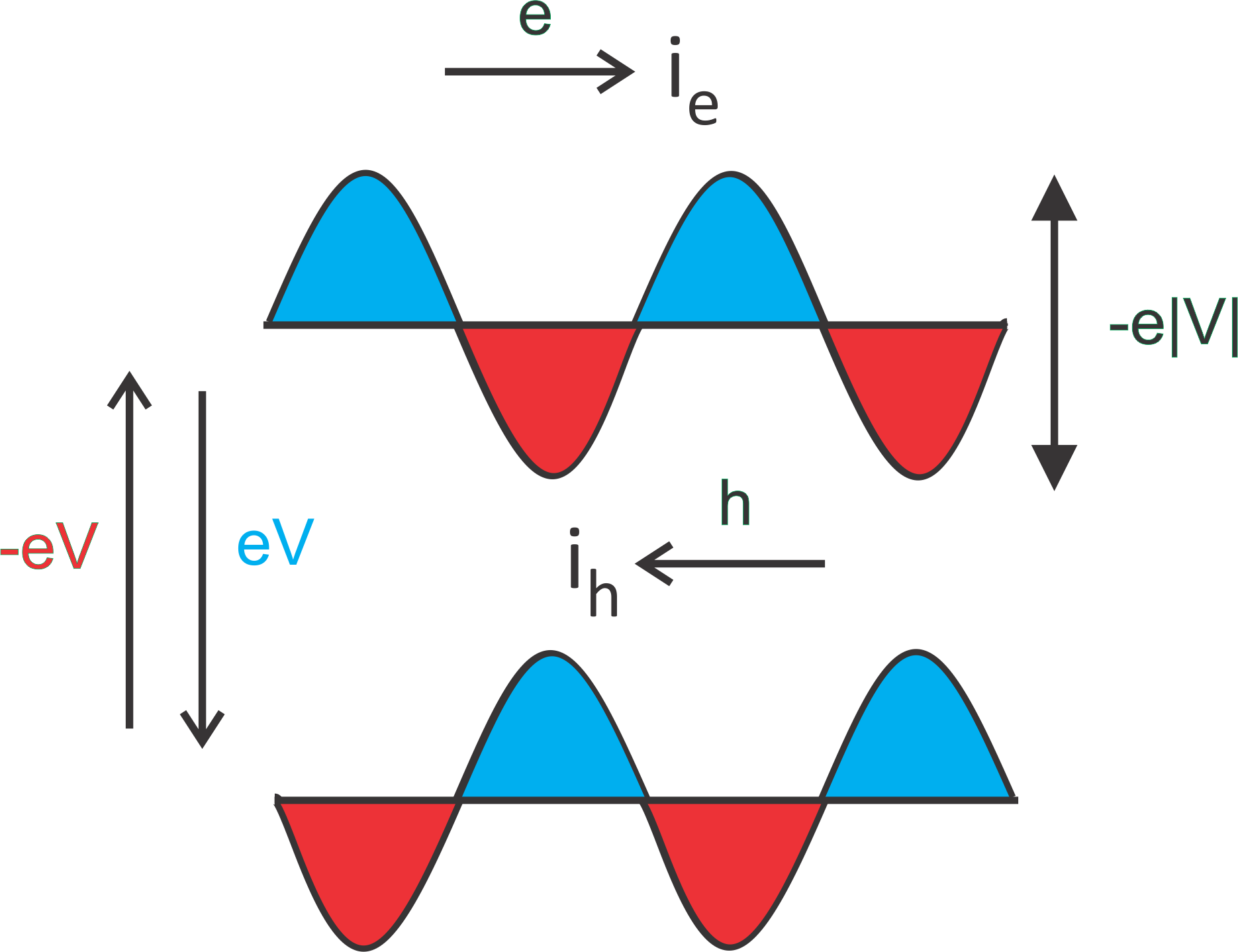}
\caption{Schematic representation of the electric current degeneracy predicted by the quantum rate theory owing to the energy degeneracy of the equivalent electrochemical capacitance of the interface $C_\mu$, which is a result of the series combination of electrostatic $C_e$ and quantum $C_q$ capacitances, where $C_e \sim C_q$. The resultant electric current $2 i_0 = C_q (dV/dt)$ is an ambipolar electric current where the electron wave flows in the opposite direction to the hole wave. An oscillatory harmonic potential perturbation of the particles in the perpendicular direction of the propagation of the wave permits to assume that $dV(t)/dt = j \omega \tilde{V}$, where $\omega$ is the frequency of perturbation over $\tilde{V} = |V|\exp \left( j \omega t \right)$ with an amplitude $|V|$ (with an energy $-e|V|$ for the electron and $e|V|$ for the hole) of the potential perturbation. Note that $i_0$ physically corresponds to a displacement electric current where there is no net transport of charge or particle mass transversal to the perturbation of the particle, which conforms with the quantum electrodynamics viewpoint of the transport of a massless Fermionic-Dirac particle, as stated in Eq.~\ref{eq:Planck-Einstein}.}
\label{fig:i-degenerancy}
\end{figure}

Two equivalent interpretations of $g_e$ degeneracy are proposed. Firstly, resonant electric currents, resulting from the combination of electrostatic and quantum capacitances, give rise to an ambipolar electric current, as depicted in Figure~\ref{fig:i-degenerancy}. This current, denoted as $i_0$, arises due to the electrolyte's role in the superposed electrostatic and quantum capacitive charging modes of molecular junctions. The equivalence of energy states $e^2/C_e$ and $e^2/C_q$ leads to an energy degeneracy that permits two electric currents contributing to the net current $i_0$ of the interface, referred to as a bipolar electric current.

In the subsequent section, the concept of $\nu$, defined by Eq.~\ref{eq:nu} and stated in terms of the conductance quantum ($G_0 = g_s e^2/h$), which has a constant value of approximately 77.5 $\mu$S, will be further elucidated.

\subsection{Conductance, Capacitance, and Quantum Mechanical Rate at Atomic and Molecular Scales}\label{sec:G0+Landauer}

It is demonstrable that $\nu$ can be expressed in terms of $G_0$~\citep{Bueno-2023-3}, a molecular-scale conductance, by accounting for the degeneracies $g_s$ and $g_e$ in the electrodynamics predicted by Eq.\ref{eq:nu}. This relationship is given by

\begin{equation}
\label{eq:nu-degenerated}
\nu = g_e g_s \left( \frac{e^2}{hC_q} \right)= g_e g_s \left( \frac{E}{h} \right) = g_e \frac{G_0}{C_q},
\end{equation}

\noindent where the atomic and molecular-scale definition of capacitance, $C = dq/dV$, is considered, in which $dq = -e dn$ and $dV = -d\mu/e$, with $d\mu$ representing the chemical potential energy associated with the perturbation of potential $dV$. The chemical potential energy per unit charge can be defined as $d\mu / dn = e^2/C_q$. For a single state $dn = 1$, this provides the capacitance of a single resonance mode, leading to $\Delta \mu = e^2/C_q = \mu_D - \mu_A$ for this single quantum channel mode of electron transport. By noting that $dE = - edV = d\mu$, it is possible to demonstrate that $C_q = e^2 \left( dn/dE \right)$, where $\left( dn/d\mu \right)$ corresponds to the density-of-states (DOS). The analysis demonstrates that $C_q$ is directly proportional to the DOS. A more detailed analysis of the significance of $C_q$ is given in the SI (Support Information) document.

The interpretation of $\nu$ at room temperature, as described by Eq.~\ref{eq:nu-degenerated}, necessitates the incorporation of statistical mechanics considerations, accounting for the thermal broadening of energy states $E$ associated with the electrodynamics. In the subsequent section, an analysis of $C_q$ and $\nu$ incorporating statistical mechanics will be explored.

\subsection{The Statistical Mechanics of the Quantum Rate}\label{sec:Therm-QR}

The thermodynamic effects contributing to the thermal broadening of $E = e^2/C_q$ are quantifiable through statistical mechanics, employing the presumption of the grand canonical ensemble~\citep{Bueno-book-2018, Bueno-2023-3, bisquert2008review}. This leads to the expression

\begin{equation}
\label{eq:Cq-thermal}
C_{q} = \left( \frac{e^2N}{k_{B}T} \right) \left[ f(1-f) \right],
\end{equation}

\noindent where $f = \left(1 + \exp \left(E / k_{B}T \right) \right)^{-1}$ represents the Fermi-Dirac distribution function, where $k_B$ is the Boltzmann constant and $T$ is the absolute temperature. Under this statistical mechanics analysis, $C_q$, an experimental observable, transforms into

\begin{equation}
\label{eq:k-finiteT}
\nu = g_e G_0 \left( \frac{k_BT}{e^2N} \right) \left[ f(1-f) \right]^{-1},
\end{equation}

\noindent where $N$ denotes the total number of states. Eq.~\ref{eq:k-finiteT} describes the overall concept of quantum rate $\nu$ for the electronic communication between quantum states at finite-temperature. The definition of $\nu$ in Eq.~\ref{eq:k-finiteT} corresponds to the electron-transfer rate constant $k$ for diffusionless electrochemical reactions~\cite{Alarcon-2021}. Eq.~\ref{eq:k-finiteT} also incorporates Marcus electron-transfer theory~\cite{Marcus-1964} as a particular setting of quantum rate $\nu$, as demonstrated in details in references~\cite{Bueno-2020, Bueno-2023-3}. Shortly, by considering the charge-transfer of a single electron in which $N$ equations to the unit and by applying the Boltzmann approximation to $f$, in which $\exp \left(E / k_{B}T \right) >> 1$ condition is used for $f = \left(1 + \exp \left(E / k_{B}T \right) \right)^{-1}$, and finally noting the definition of $G_0 = g_s e^2/h$, Eq.~\ref{eq:k-finiteT} provides a $\nu \propto \left( k_B T/h \right) \exp \left( -E / k_{B}T \right)$ from which Marcus defined $E = \left( eV + \lambda /  \right)^2/4\lambda$, with $\lambda$ as the reorganization energy of the solvent. 

Additionally, at the Fermi level of a junction, where $f = 1/2$, and with a single electron transmittance in an ideal adiabatic mode, $\nu$ of Eq.~\ref{eq:k-finiteT} simplifies to $\nu = k_BT / h$, as anticipated, demonstrating the adherence to quantum mechanical principles at room temperature and the dynamics predicted by the Planck-Einstein relationship for $E = k_BT$. This analysis implies that the electron dynamics between push-pull molecular states and electrodes adhere to Dirac (relativistic) rather than Schr\"odinger (non-relativistic) equations, as will be elucidated in this manuscript, showcasing that the communication rate is quantized at room temperature, with a value of $G_0/C_q = g_sE/h$ for a perfectly adiabatic quantum channel transmittance setting.

\begin{figure}[h]
\centering
\includegraphics[width=5cm]{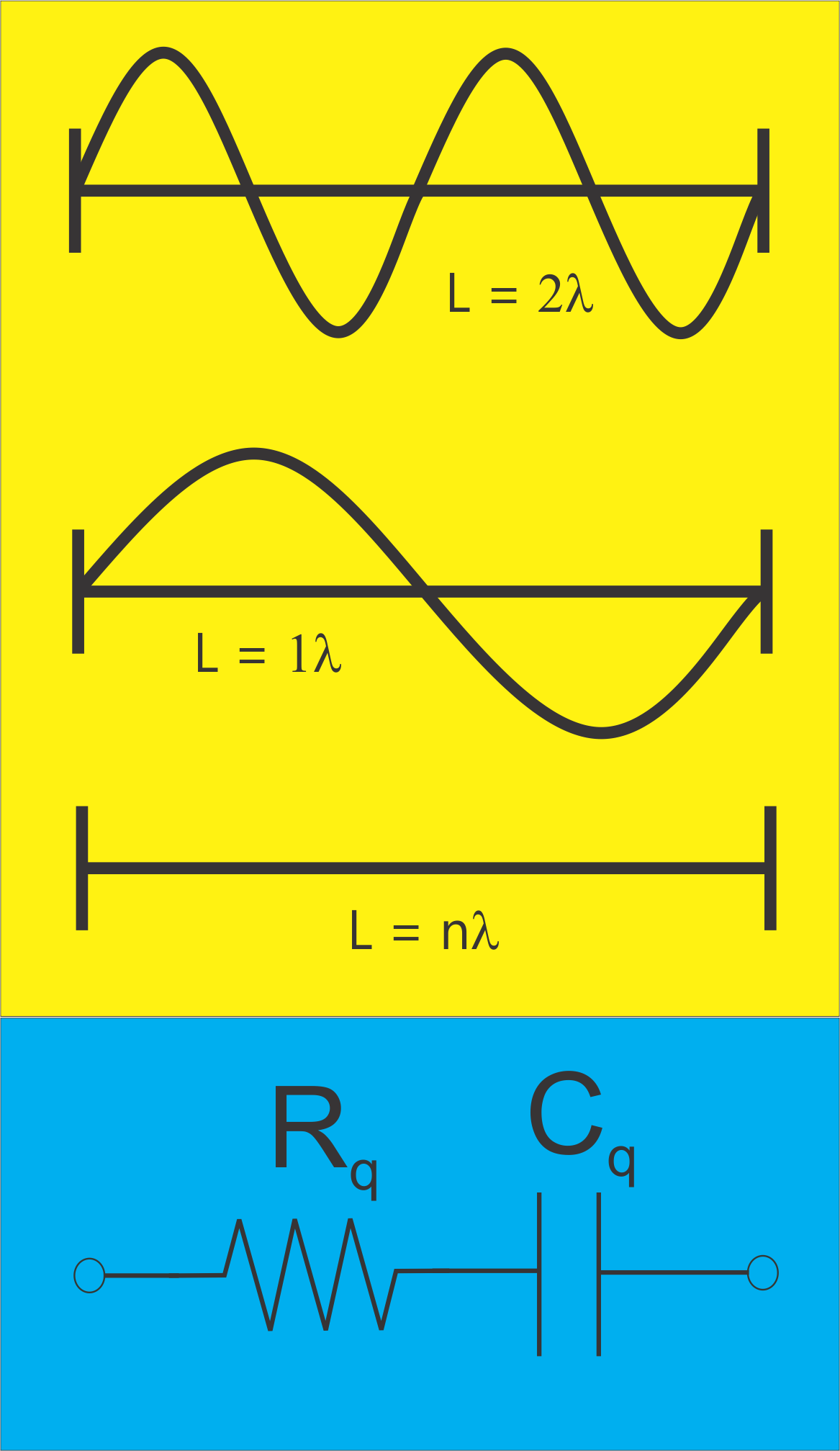}
\caption{According to the quantum rate theory within a $\pi$ molecular bridge of length $L$ separating $D$ and $A$ of push-pull molecular structure, there can be different $n$ wavelength quantum modes $\lambda$ for an adiabatic intra-molecular electron transfer dynamics to occurs. Independently of the quantum modes of electron transport, the quantum electrodynamics can be studied, according to dynamics predicted by Eq.~\ref{eq:Complex-Cq}, by measuring $\nu = 1/\tau = G_0/C_q$, where $\tau = R_q C_q$ with $R_q = 1/G_0$. In practical experimental terms, $R_q \sim$ 12.9 k$\Omega$ corresponds to the total series resistance of the molecular junction, encompassing $R_e$ and $R_c$ state for the resistances of electrolyte and contact, respectively.}
\label{fig:Lmode+circuit}
\end{figure}

Owing to $\nu \propto G_0/C_q = 1/R_q C_q = 1/\tau$, where $R_q = 1/G_0\sim$ 12.9 k$\Omega$ represents the resistance quantum and $\tau = R_q C_q$ signifies the quantum RC characteristic time of the molecular junction's electrodynamics, we can model the electrodynamics using quantum RC-circuit dynamics. This also permits us to utilize the electric impedance of the interface as a valuable quantum mechanical spectroscopic method, as will be demonstrated in the subsequent section.

\subsection{Equivalent Circuit and Spectroscopic Analysis of the Quantum Rate Dynamics}\label{sec:QRS-circuit}

Within the framework of the quantum rate theory~\citep{Bueno-2023-1, Bueno-2023-2, Bueno-2023-3}, the quantum RC dynamics of push-pull molecular junctions can be examined using impedance-derived capacitance spectroscopy, wherein a complex capacitance~\citep{Bueno-2023-3, buttiker1993mesoscopic} can be measured and analyzed. This complex capacitance is represented by

\begin{equation}
\label{eq:Complex-Cq}
C^{*}(\omega) = \frac{C_q^0}{1 + j\omega \tau} \approx C_q^0 \left(1 - j\omega \tau \right),
\end{equation}

\noindent where $C_q^0$ in Eq.~\ref{eq:Complex-Cq} denotes the equilibrium capacitance, attained at $\omega \rightarrow 0$, where the imaginary component of the complex capacitance becomes negligible.

A spectroscopic examination based on Eq.\ref{eq:Complex-Cq} allows for the construction of useful DOS \textit{versus} energy spectra, a technique referred to as quantum rate (QR) spectroscopy~\citep{pinzon2023quantum, Lopes-2023}. For instance, the analysis of QR spectra, utilizing Eq.~\ref{eq:Complex-Cq}, has been successfully employed to ascertain the electronic structure of graphene~\cite{Lopes-2023} and quantum dots~\cite{pinzon2023quantum}. In the present study, we scrutinize complex capacitive spectra based on Eq.~\ref{eq:Complex-Cq} for push-pull molecular junctions, enabling us to investigate the applicability of the QR theory in probing the relativistic quantum electrodynamics of organic semiconducting junctions.

\section{Experimental }\label{sec:experimental}

\subsection{Chemical Reagents and Solutions}

Commercially available ethyl 2-azidoacetate (N$_3$CH$_2$CO$_2$C$_2$H$_5$), sodium ethoxide, (NaOC$_2$H$_5$), sodium hydroxide (NaOH), ethanol, xylene, tetrabutylammonium perchlorate (TBAClO$_4$), dimethylformamide (DMF) and cysteamine were purchased from Sigma–Aldrich. \textit{N},\textit{N'}-Diisopropylcarbodiimide (DIC), and 1-hydroxybenzotriazole hydrate (HOBT) were purchased from Chess GmbH. All chemicals were used as received, without further purification. Milli-Q water was used for all the experiments.

\subsection{Instrumentation}

Melting points were determined using a Stuart SMP3 melting point apparatus. TLC analyses were conducted on silica plates pre-coated with a 0.25 mm layer (Merck Fertigplatten Kieselgel 60F$_{254}$), and spots were visualized under UV light. NMR spectra were acquired on a 250 MHz Bruker AC 250, 300 MHz Varian Unity Plus Spectrometer, and Bruker Avance III 400, operating at frequencies of 62.9 MHz for $^1$H and 100.6 MHz for $^{13}$C. Chemical shift values ($\delta$ relative to TMS) are provided in parts per million (ppm), with the solvent peak serving as an internal reference at 25 $^\circ$C.

\subsection{Synthesis and characterization of 4\textit{H}-thieno[3,2-\textit{b}]pyrrole-5-carboxylic acid (\textbf{1, TPC})}

\begin{figure*}
    \centering
    \includegraphics[width=17cm]{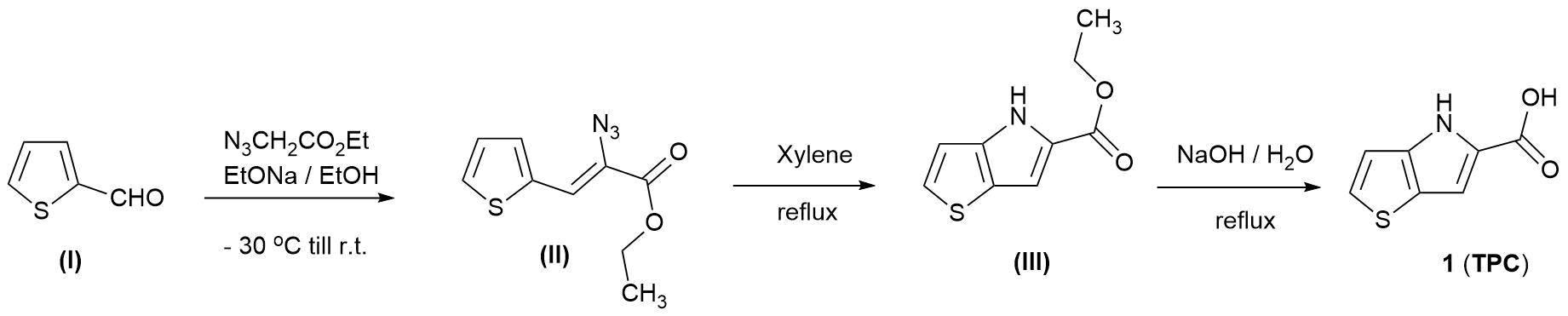}
    \caption{Schematic depiction of the sequence of synthesis of thienylpyrrole derivative (1, TPC) applied for constructing the molecular junctions studied in the present work.}
    \label{fig:synthesis}
\end{figure*}

The thienylpyrrole derivative (\textbf{1, TPC}) was synthesized via condensation of thiophene-2-carbaldehyde (\textbf{I}) and ethyl azidoacetate in the presence of sodium ethoxide and ethanol at -30°C, followed by warming to room temperature, yielding ethyl 2-azido-3-(thiophen-2-yl)acrylate (\textbf{II}). Subsequent reflux of azide (\textbf{II}) in toluene afforded the condensed system ethyl 4\textit{H}-thieno[3,2-\textit{b}]pyrrole-5-carboxylate (\textbf{III})\cite{keener1968synthesis}. Hydrolysis of compound \textbf{III} using NaOH aqueous solution yielded the corresponding 4\textit{H}-thieno[3,2-\textit{b}]pyrrole-5-carboxylic acid (\textbf{1, TPC}), as depicted in Figure~\ref{fig:synthesis}. All compounds underwent thorough characterization, and their structures were confirmed using standard spectroscopic techniques. Compound \textbf{1} was obtained for the first time with an 85$\%$ yield using a different synthetic approach \cite{welch1999improved}.

\subsection{Preparation of \textbf{TPC} Assembly}

Gold electrodes (Au electrodes) were fabricated by radio frequency sputtering, depositing titanium and gold sequentially onto silicon oxide wafers measuring 0.8 x 0.5 cm, resulting in electrodes with an average geometric area of approximately 0.04 cm$^{-2}$. These Au electrodes underwent a cleaning process involving sequential immersion in solvents: isopropyl alcohol, acetone, isopropyl alcohol, and ultrapure water, each followed by sonication for 12 minutes. Subsequently, the surfaces were dried with N$_2$ and cleaned under ultraviolet/ozone for 30 minutes. The pretreated Au electrodes were then immersed in an aqueous solution containing 50 mmol L$^{-1}$ of cysteamine for 16 hours at room temperature, facilitating the formation of a cysteamine monolayer over the Au surface. Following this step, the electrodes were rinsed thoroughly with ultrapure water.
Next, the cysteamine monolayer-modified Au electrodes were immersed in a solution containing 1.0 mmol L$^{-1}$ of 4\textit{H}-thieno[3,2-\textit{b}]pyrrole-5-carboxylic acid (\textbf{1, TPC}) in dimethylformamide (DMF), supplemented with 50 mmol L$^{-1}$ of diisopropylcarbodiimide (DIC) and 50 mmol L$^{-1}$ of 1-hydroxybenzotriazole (HOBT). This incubation step occurred in the dark for 4 hours. Finally, the \textbf{TPC} molecule-modified Au electrodes were rinsed again with ultrapure water prior to any measurements being conducted.

\subsection{Electrochemical Measurements}

Electrochemical impedance spectroscopy (EIS) measurements were conducted using a portable potentiostat (PalmSens4) equipped with a frequency response analyzer (FRA) in a three-electrode setup. The working electrode was the \textbf{TPC} molecule-modified gold electrode, while a platinum mesh served as the counter electrode. All electrochemical measurements were performed in a 20 mmol L$^{-1}$ solution of tetrabutylammonium perchlorate (TBAClO$_4$) in dimethylformamide (DMF). 

EIS spectra were recorded at a fixed bias potential corresponding to the open circuit potential (OCP), which in this study was $-$0.62 V. A root mean square (RMS) amplitude potential perturbation of 10 mV was applied, with a frequency ranging from 1 MHz to 0.05 Hz. From the raw EIS data, the real ($C'$) and imaginary ($C''$) capacitive components were obtained using the relationship $Z^* (\omega)= 1 / j\omega C^* (\omega)$, where $Z^* (\omega)$ and $C^* (\omega)$ are the impedance and capacitance complex functions, respectively. Here, $\omega$ represents the perturbing angular frequency and $j$ refers to the imaginary unit $\sqrt{-1}$. The complex $C_q^(\omega)$, consistent with Eq.~\ref{eq:Complex-Cq}, was expressed as $C_q^ (\omega)= C' + jC''$, where $C'$ and $C''$ were derived from the real ($Z'$) and imaginary ($Z''$) components of the measured complex impedance function $Z^*(\omega)= Z' + jZ''$. Note that $|Z|=\sqrt{(Z')^2 + (Z'')^2}$ represents the modulus of impedance. The diameter of the semicircle observed in the Nyquist capacitive diagram, where the $C''$ component is minimal, corresponds to a frequency denoted as the equilibrium frequency ($\omega_0$). This frequency $\omega_0$ characterizes the charge equilibrium state and is determined for $\omega \rightarrow 0$ in Eq.~\ref{eq:Complex-Cq}. It allows access to the DOS function as a function of the junction potential (or electrode potential), which in this study ranged from $-$2.4 to $-$0.5 V \textit{versus} a platinum pseudo-reference electrode.

\section{Results and Discussions}\label{sec:Results}

\begin{figure}[h]
\centering
\includegraphics[height=5.5cm]{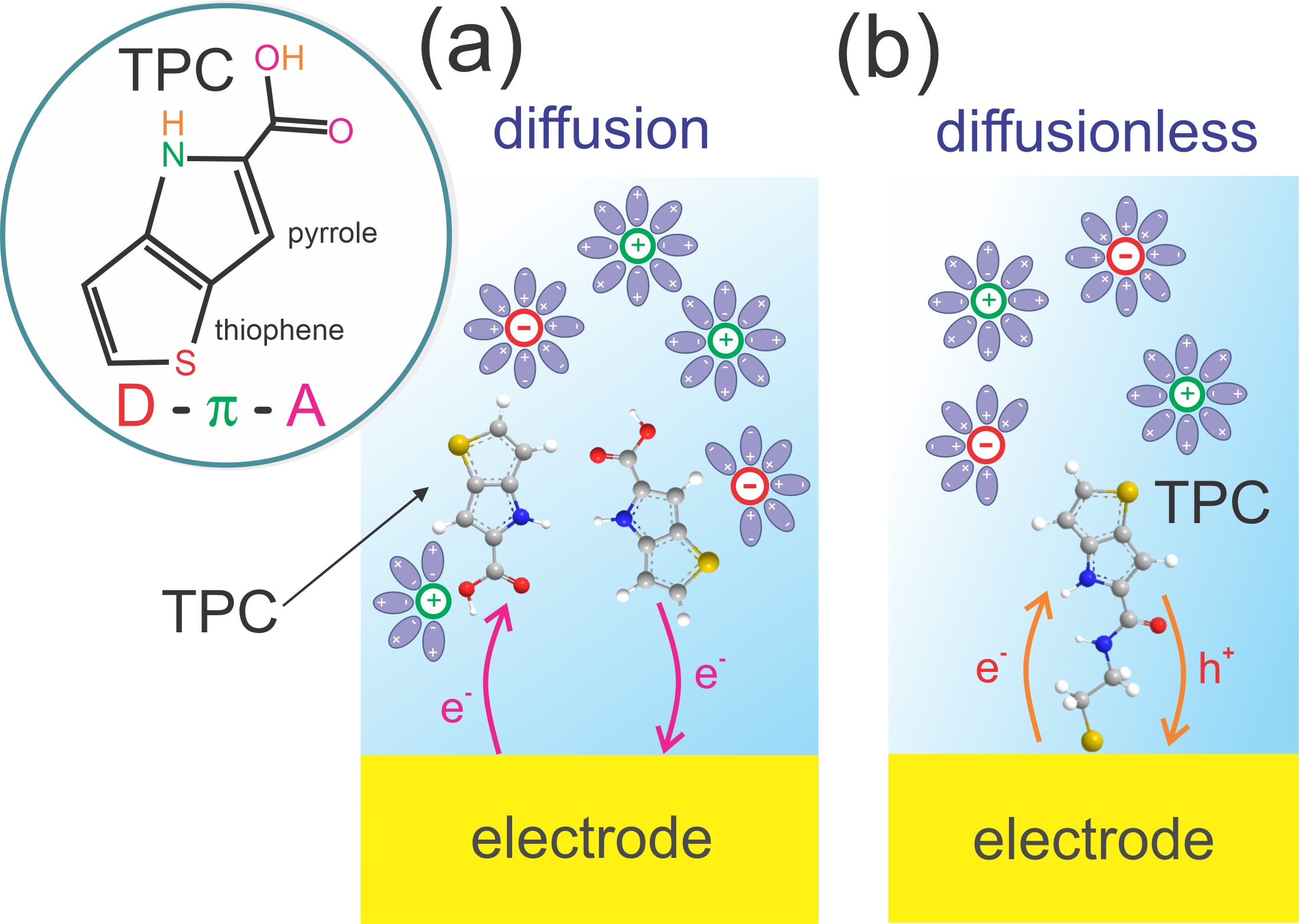}
\caption{Schematic representation of two electrochemical setups for studying the electronic structure of push-pull heterocyclic molecules, focusing on the structure of 4\textit{H}-thieno[3,2-\textit{b}]pyrrole-5-carboxylic acid (\textbf{1, TPC}) molecule in two distinct electrode configurations. (a) This depicts the conventional electrochemical setup, where electron transfer (ET) between the electrode and \textbf{TPC} molecules, freely existing in an electrolytic bulk environment, occurs under diffusion-limited conditions. (b) Illustrates the electrochemical configuration investigated in this study, where \textbf{TPC} molecules are covalently bonded to electrodes, forming a single-contact molecular junction structure. The supporting electrolyte provides an appropriate electric-field screening condition to explore the electronic properties of the \textbf{TPC} molecular ensemble by quantum rate theory. The analysis's alignment with quantum rate theory confirms the quantum electrodynamics governing the interaction of the electrode with the quantum states of the \textbf{TPC} molecules.}
\label{fig:EC-settings}
\end{figure}

As noted in section~\ref{sec:introduction}, electrochemistry is frequently used to characterize the electronic properties of push-pull heterocyclic compounds. However, the use of a current-voltage pattern provided by cyclic voltammetric methods is applied to access the electronic properties of push-pull molecules free in an electrolyte medium, as depicted in Figure~\ref{fig:EC-settings}\textit{a}. As is the case for redox reactions, this experimental setting, in which the investigated $D$ and $A$ moieties are freely in the electrolyte environment, implies that there is a classical diffusional control of the electron transport from $D$ species to the electrode or the electrode to $A$ species, with the electron transport dynamics controlled by mass diffusion of $D$ and $A$ species from/to the surface of the electrode. In other words, in this setting of electrochemical investigation, the electron transport has a charge transfer kinetics controlled by mass diffusion. Hence, a diffusional control of the electron transport implies that there is no associated quantum capacitance or quantum RC dynamics, as discussed in the context of Eq.~\ref{eq:Complex-Cq} and, consequently, for this type of electrochemical experimental settings, there is no quantum electrodynamics within a quantum rate approach to be studied.

\begin{figure*}[h]
\centering
\includegraphics[height=7.0cm]{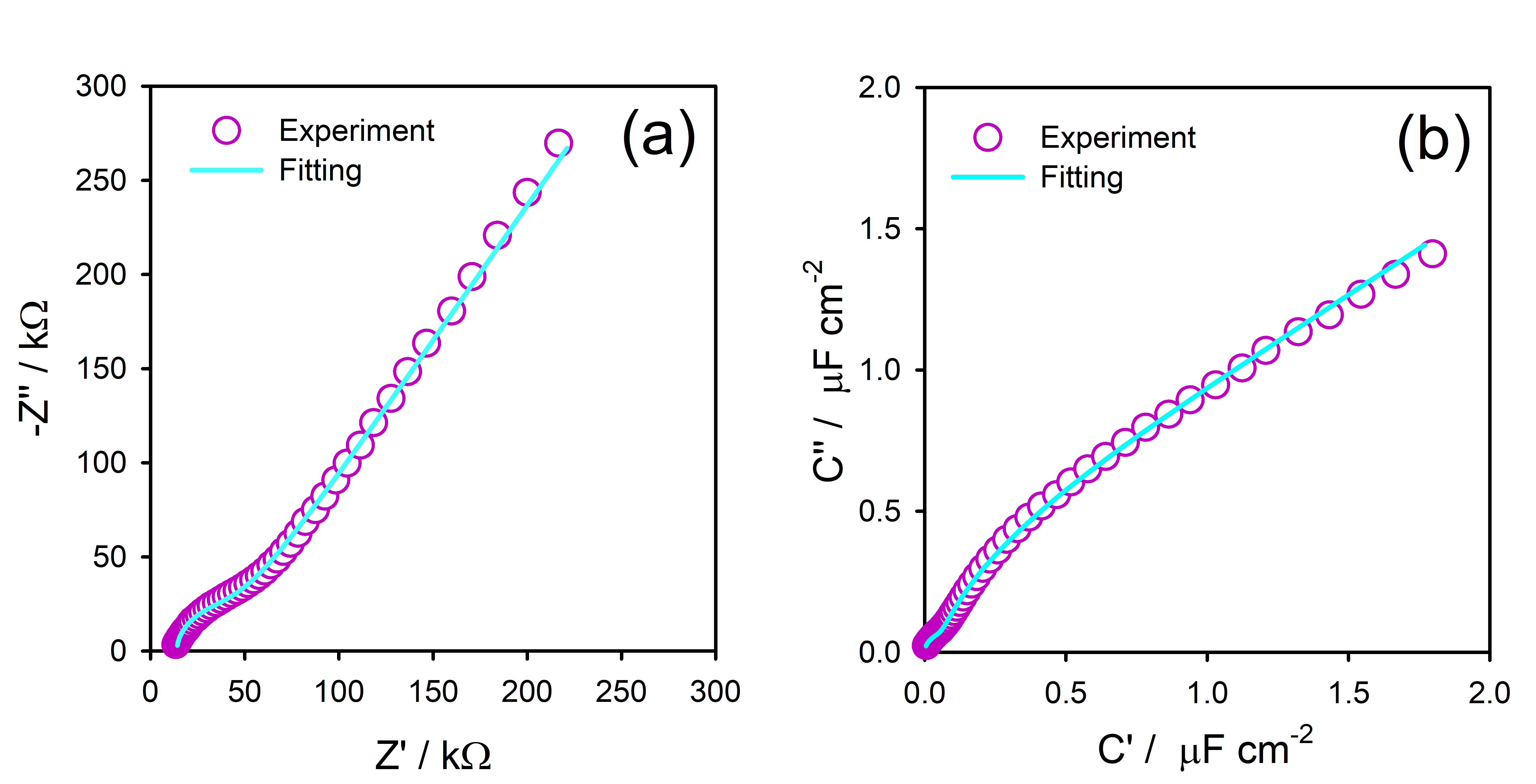}
\caption{(a) Impedance and (b) capacitance Nyquist diagrams obtained from the impedance spectroscopic measurements of \textbf{TPC} free in an electrolyte environment and in contact with a conducting electrode. The electrolyte, in which 1 mmol L$^{-1}$ \textbf{TPC}, was added to DMF with 20 mmol L$^{-1}$ of TBAClO$_4$. These spectra were collected at a potential bias of $-$0.65 V, corresponding to the first reduction peak in the cyclic voltammogram shown in Figure~S1. These spectra were well-fitted to the Randle equivalent circuit model illustrated in Figure~\ref{fig:circuit-comparisons}\textit{a}, confirming the governance of ET kinetics of the reactants to the electrode by diffusion.}
\label{fig:EIS-free}
\end{figure*}

The above analysis can be confirmed by investigating the current-voltage (CV) pattern, as depicted in Figure~S1 of the SI document, obtained by placing a gold working electrode, used as a probe, in direct contact with \textbf{TPC} moieties diluted in an electrolyte in a concentration of 1 mmol L$^{-1}$. From the CV pattern, a potential energy separation, $E_g$ $\sim$ 2.08 eV between the first oxidation and reduction current peaks is measured with the potential difference of these peaks not owing be to a resonant dynamics between $D$ and $A$ energy states of \textbf{TPC} communicating with the electrode, which is confirmed by conducting an impedance measurement at first reduction peak of the CV pattern. Impedance and capacitance spectra at this peak, corresponding to a potential of $-$0.65 V in the CV, are shown in Figure~\ref{fig:EIS-free}. The impedance spectrum (Figure~\ref{fig:EIS-free}\textit{a}) is typical of a kinetically diffusion-limited process, and the measured electrochemical current in the CV peak is confirmed to be owing to an irreversible reduction of the \textbf{TPC} compounds, rather than a resonant electric current.

Also, the Nyquist capacitance spectra measured at the reduction potential of $-$0.65 V \textit{versus} platinum (as pseudo-reference potential), shown in Figure~\ref{fig:EIS-free}\textit{b}, confirmed the above interpretation of the CV pattern. The analysis of this capacitive spectrum is consistent with the irreversible transfer of the electrons from the electrode to the $A$ state (carboxylic acid group, see inset of Figure~\ref{fig:EC-settings}\textit{a}) of the \textbf{TPC} molecule, as expected for the measurement conducted at the $-$0.65 V.

\begin{figure}[h]
\centering
\includegraphics[height=4.7cm]{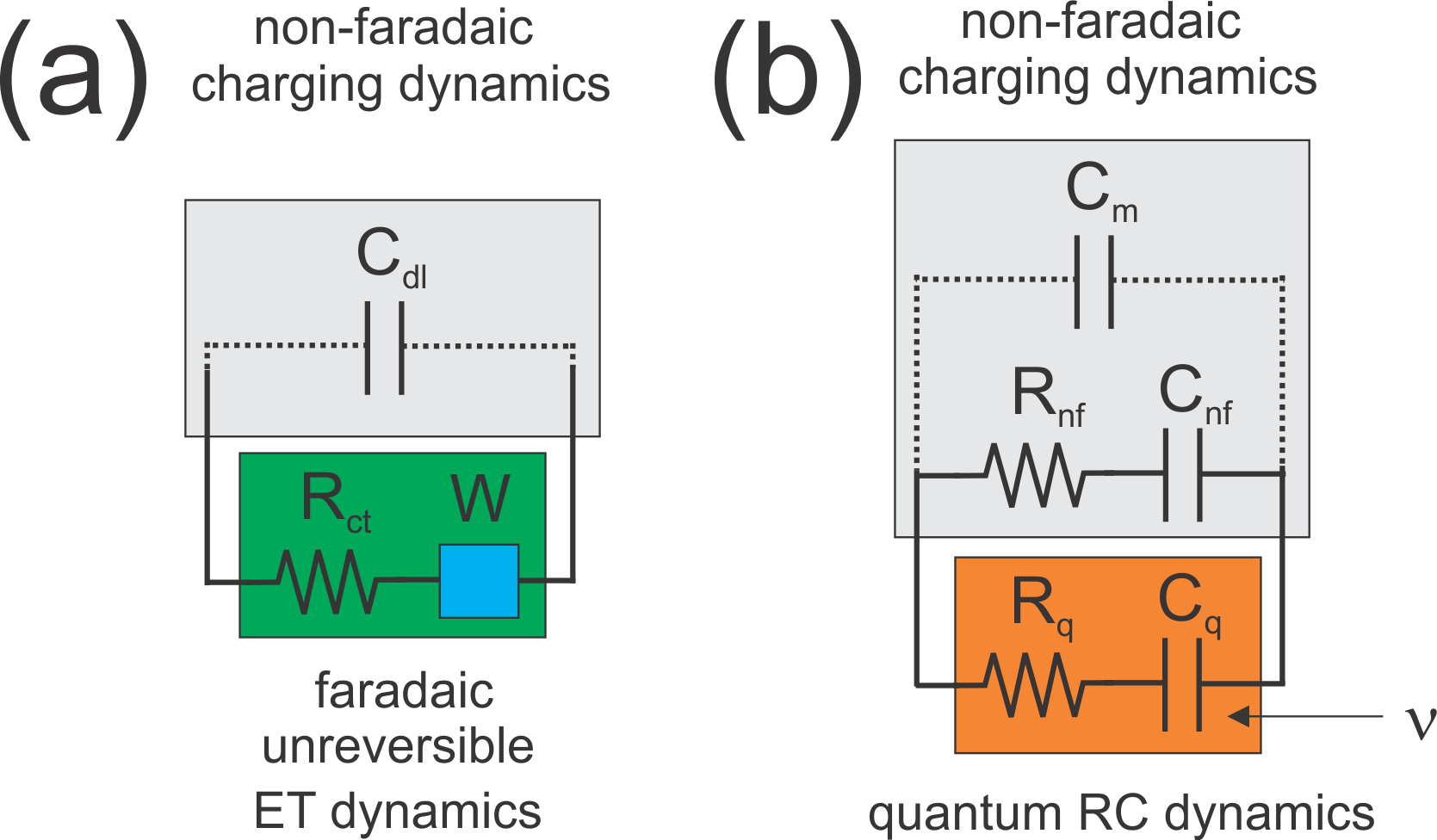}
\caption{(a) The traditional Randle equivalent circuit models the direct electric contact of electro-active molecules or compounds, whether undergoing electrochemical reactions or not, within an electrolyte medium, with a conducting electron reservoir (a probe electrode). The series combination of charge transfer resistance $R_{ct}$ and Warburg $W$ circuit elements in the equivalent circuit branch (depicted in green) signifies electron transfer kinetics governed by diffusion. The presence of the $W$ element accounts for electrodynamics controlled by the transport of mass of reacting species toward the electrode. The parallel capacitance $C_{dl}$ corresponds to double-layer capacitance.
(b) This equivalent circuit models the electrodynamics of molecularly modified electrodes within an electrolyte environment. The non-faradaic charging dynamics (illustrated in gray) refers to dynamics that do not follow quantum RC dynamics (depicted in red), which involve a series combination of quantum resistive $R_q$ and capacitive $C_q$ elements. The quantum RC dynamics obey the quantum rate dynamics predicted by the quantum rate model, where $\tau = R_q C_q$ corresponds directly to $\nu = 1/\tau \propto e^2/hC_q$ rate.}
\label{fig:circuit-comparisons}
\end{figure}

Accordingly, both impedance and capacitive Nyquist diagrams of Figure~\ref{fig:EIS-free} confirmed a diffusion-controlled electron transfer process kinetically limited by the transport of \textbf{TPC} molecules from the electrolyte bulk medium to the surface of the electrode, and this diffusion-limited ET electric current is well-modeled using a Randle circuit, as shown (see Figure~\ref{fig:circuit-comparisons}\textit{a}). Hence, the ET between the electrode and \textbf{TPC} molecules free in solution is governed by a series combination of charge transfer resistance $R_{ct}$ and Warburg ($W$) impedance circuit elements, as indicated in the green box of Figure~\ref{fig:circuit-comparisons}\textit{a}. The Warburg impedance element accounts for a diffusion-controlled mass process within a diffusion length region in the surface of the electrode.

The fitting of the impedance and capacitance spectra, shown (blue line) in Figure~\ref{fig:EIS-free}, to the Randle circuit of Figure~\ref{fig:circuit-comparisons}\textit{a} is quite good (with a confidence of 99\%) and confirms the diffusion-controlled characteristics of this form of electrochemically evaluating the \textbf{TPC} push-pull electronic structure. For the above analysis, an almost three times larger value than the quantum resistance theoretical limit $\sim$ 12.9 k$\Omega$ was obtained, i.e. $R_{ct} \sim$ 42.46 k$\Omega$.

An additional electrochemical way of characterizing push-pull molecular electronic junctions that complies with quantum rate theory and opens a new possible interpretation of the electrodynamics of molecular junctions in an electrochemical medium, will be evaluated in the next section.

\subsection{Resonant Molecular Junctions at Electrochemical Environments}\label{sec:RMJ+EC}

In contrast to the electrochemical setting of measuring the electronic properties of push-pull molecular junctions, discussed in the previous section, another way of characterizing the electronic properties of \textbf{TPC} heterocyclic molecules that correlates with the quantum rate theory will be discussed here. As depicted in Figure~\ref{fig:EC-settings}\textit{b}, \textbf{TPC} molecules covalently anchored to a probe conductive electrode, from which the molecular ensemble, is investigated, corresponding to a molecular junction with similar quantum electrodynamics to that measured in redox-active molecular junctions, in which explicitly there are electrochemical reactions and hence electron transfer (ET) undergoing in the junction. However, there are two main differences between push-pull and redox-active junctions: (i) in push-pull junctions, there is no ET process between the molecule and the electrode, and (ii) the quantum conductance in the junctions follows adiabatic electrodynamics.

Accordingly, if an experimental setting, as that illustrated in Figure~\ref{fig:EC-settings}\textit{b}, is used for obtaining information on the \textbf{TPC} organic/molecular electronic junction, the communication of quantum resonant states of the \textbf{TPC} junction with the electrode occurs (i) without ET and (ii) adiabatically, as confirmed by studying the lower frequency region of the impedance/capacitance spectrum measured in this electrochemical setting. The analysis is different from that depicted in Figure~\ref{fig:EC-settings}\textit{a} and in the experimental setting of Figure~\ref{fig:EC-settings}\textit{b} push-pull molecules are covalently attached to the electrode, with the presence of $C_q$ in place of $W$ circuit element. It implies that, from a physics viewpoint, there is a diffusionless electron communication of the resonant quantum states of \textbf{TPC} molecules with the electrode.

Furthermore, the presence of a $C_q$ term in the \textbf{TPC}-electrode junction implies that the DOS of \textbf{TPC} resonant states is accessible from a potential perturbation in the electrode and the communication of these states with those of the electrode follows quantum electrodynamics, as predicted by the quantum rate theory and discussed in section~\ref{sec:introduction}. That the resonant electron communication of the molecules with the electrode involves an electric (resonant) current will now be demonstrated. The physical nature of this current implies an ambipolar electric current even though there is no associated redox reaction nor an effective ET reaction occurring in this type of molecular junction. Although there is no redox reaction, the frequency (or the comunication dynamics) obeys relativistic quantum mechanical rate characteristics, as predicted by Eq.~\ref{eq:k-finiteT}, in agreement with the quantum rate theory premises. Accordingly, in kinetic terminology, the electron transmittance of wave communication occurs in diffusionless controlled dynamics (Figure~\ref{fig:EC-settings}\textit{b}), equivalent to that discussed in references~\cite{Bueno-2023-3, Alarcon-2021, Sanchez-2022-1} for electrochemical reactions, albeit no ET is effectively occurring.

The resonant electric current (referred to in physics as a displacement electric current) allows us to investigate the quantum electrodynamics of push-pull molecular junctions, measured by imposing a time-dependent perturbation. As the \textbf{TPC} junction is formed by the covalent attachment of \textbf{TPC} molecules over a conducting electrode, as illustrated in Figure~\ref{fig:EC-settings}\textit{b}, there is $\pi$-type bridge group that connects $D$ or $A$ to the electrode. Therefore, there is a $D/A-\pi$ resonance communication with the electrode, in which the $A$ group of the pristine $D-\pi-A$ electronic structure of the \textbf{TPC} compound, covalent attachment to the electrode, as detailed in section~\ref{sec:experimental}, permits to the remaining $D$ state to perform either as $D$ or $A$ states for the energy level of the electrode (for more details, refers to Figures~\ref{fig:D-pi-A-scheme} and~\ref{fig:electrode-DA}). The attachment of \textbf{TPC} molecules to the electrode was attained by chemically modifying the carboxylic acid $A$ group of the \textbf{TPC} molecules, allowing the chemical attachment of this $A$ group to the electrode through amine groups present in a previously assembled cysteamine molecular film over the gold electrode.

\begin{figure}[h]
\centering  
\includegraphics[height=7cm]{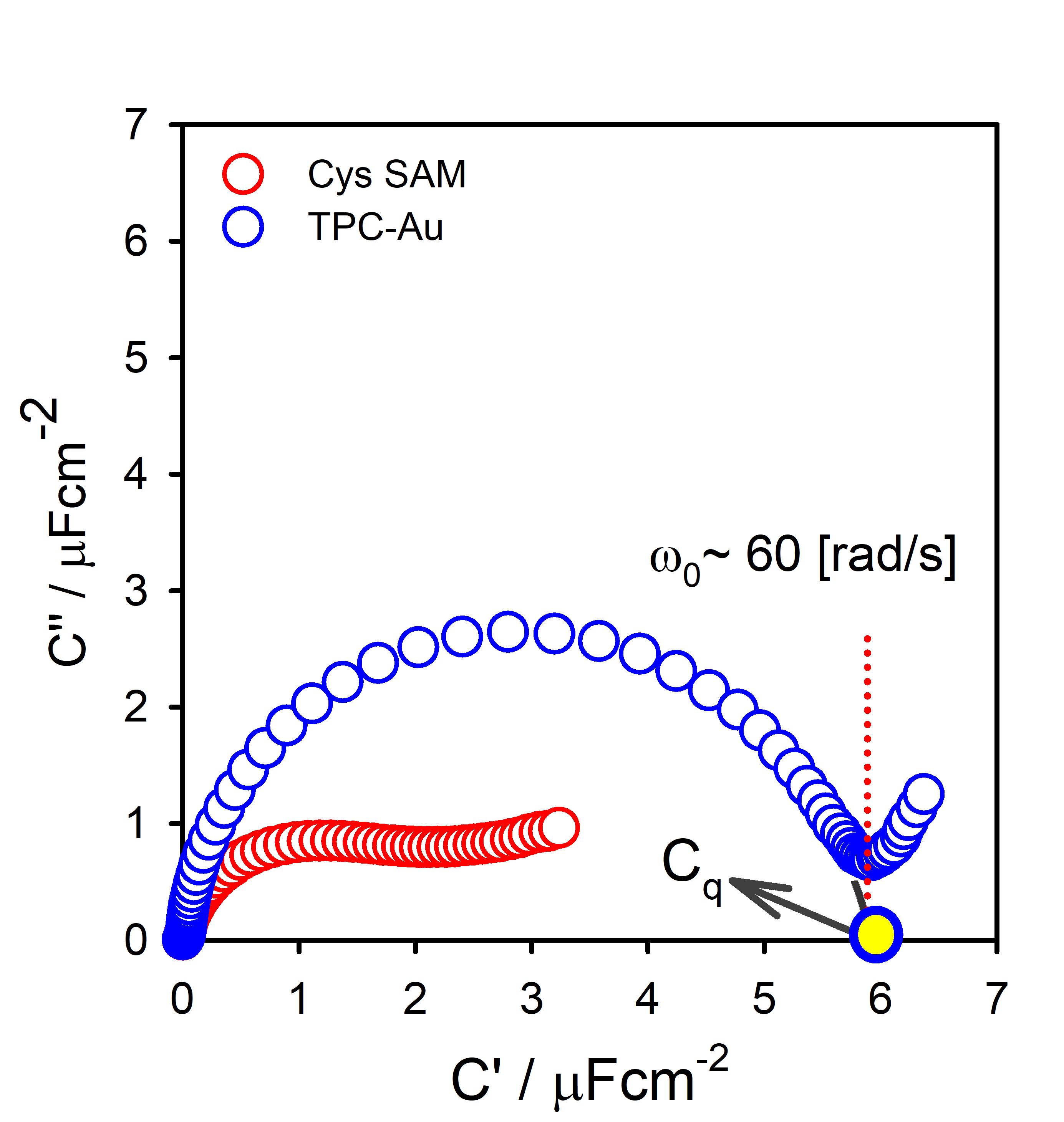}
\caption{Nyquist capacitive diagrams measured at the OCP for the cysteamine (denoted in red as Cys SAM) and \textbf{TPC} (indicated in blue as \textbf{TPC})-Au junctions in 20 mmol L$^{-1}$ of TBAClO$_4$. The \textbf{TPC} junction comprises \textbf{TPC} molecules chemically attached to the surface of the electrode employing the cysteamine film. The $C_q$ of the interfaces is accessible as the measurement of the real component of the complex capacitance obtained as the diameter of the Nyquist capacitive diagram. The corresponding frequency for this value of $C_q$ corresponds to the charge equilibrium frequency indicated as $\omega_0$.}
\label{fig:Nyquist-OCP}
\end{figure}

Figure~\ref{fig:Nyquist-OCP} shows the Nyquist capacitive spectrum of the \textbf{TPC} molecular junction (blue circles) recorded at the open circuit potential (OCP) and compared with the Nyquist capacitive diagram recorded only with the cysteamine film (red circles), before the chemical attachment of the \textbf{TPC} organic structure to the interface containing the cysteamine film. The differences concerning the nature of the electric current $i = C_q s$ are evident in both molecular films, despite an equivalent magnitude of the electric currents, as shown in Figure~\ref{fig:EIS+DOS-FLevel+QC}\textit{b}, for each of these interfaces. The quantum RC relaxation characteristics are evident in the semi-circle shape of the junction containing \textbf{TPC} anchored moieties with the characteristic resonance of the electric current associated with $C_q$ confirmed in Figure~\ref{fig:EIS+DOS-FLevel+QC}\textit{b}, where now the prevailing resonant electric current $i_0 = C_q s$ measured at the Fermi level of the molecular junction is at least two orders of magnitude higher than that observed in the OCP.

Note that the data shown in Figure~\ref{fig:EIS+DOS-FLevel+QC} were measured by firstly constructing the electronic DOS curve of the interface, in agreement to the theory introduced within Eq.~\ref{eq:Cq-thermal}, corresponding to measuring $C_q \propto$ DOS at the room temperature and ambient electrolyte conditions. This room and electrolyte ambient implies that $C_q$ is related to a thermal broadened energy $E = e^2/C_q$ state of the interface, as theoretically introduced in section~\ref{sec:Therm-QR}. The DOS curve, shown in Figure~\ref{fig:EIS+DOS-FLevel+QC}\textit{b}, was measured by considering a frequency of $\omega_0 \sim$ 60 rad/s, as indicated in Figure~\ref{fig:Nyquist-OCP}, in agreement to the quantum RC dynamics predicted by Eq.~\ref{eq:Complex-Cq}. In other words, the measurement of $C_q$ at different potentials of the electrode for a fixed $\omega_0 \sim$ 60 rad/s frequency allowed us to construct the DOS curves for both interfaces, as shown in blue and red dots in Figure~\ref{fig:EIS+DOS-FLevel+QC}\textit{b}.

Note that the $\omega_0 \sim$ 60 rad/s corresponds to the lower possible frequency for Eq.~\ref{eq:Complex-Cq}, which theoretically corresponds to a DC electric measurement. However, for this lower-energy quantum RC dynamics, there is no physical correspondence to DC measurements or $\omega_0$ equating to zero. The latter is owing to the quantum meaning of the conductance, which does not leave a DC-zeroed resistive possibility of interpreting the experiment. The minimum value of $\omega_0$ is equivalent to $\omega_0 = 1/\tau$ that corresponds to a maximum rate owing to the minimum value of $R_q = g_s e^2/h \sim$ 12.9 k$\Omega$ allowed for the quantum communication between two electronic states in the interface, as better discussed in the context of the quantum rate theory applied to the case of the electrodynamics within graphene~\citep{Bueno-2022}.

From the DOS pattern obtained for cysteamine (red circles) and \textbf{TPC} (blue circles) junctions, as indicated  
in Figure~\ref{fig:EIS+DOS-FLevel+QC}\textit{b}, it is evident that only the \textbf{TPC} molecular junction contains aromatic states resonating with the electrode, thus presenting a significant resonant DOS contribution within a $i_0$ ambipolar type of displacement electric current. Astonishing is the fact that albeit there is no redox reaction or electrochemical faradaic states in the interface, there is a significant pseudo-capacitive contribution at the Fermi level ($-$1.43 V \textit{versus} platinum pseudo-potential reference), corresponding to a capacitance value $\sim$ 225 $\mu$F cm$^{-2}$. This maximum capacitance value is measurable at the peak of the $C_q = i_0/s$ \textit{versus} potential curve of Figure~\ref{fig:EIS+DOS-FLevel+QC}\textit{b}. This value of capacitance is confirmed by measuring the Nyquist capacitive diagram at the Fermi level of the junction, obtained as the diameter value of the semi-circle of the Nyquist capacitive diagram, as shown in Figure~\ref{fig:EIS+DOS-FLevel+QC}\textit{a}.

The relativistic room-temperature quantum electrodynamics of \textbf{TPC}-electrode junctions, is demonstrated in the next section.

\subsection{Quantum Electrodynamics of Molecular Junction at Room temperature}\label{sec:QED-roomT}

\begin{figure*}[h]
\centering
\includegraphics[height=7.5cm]{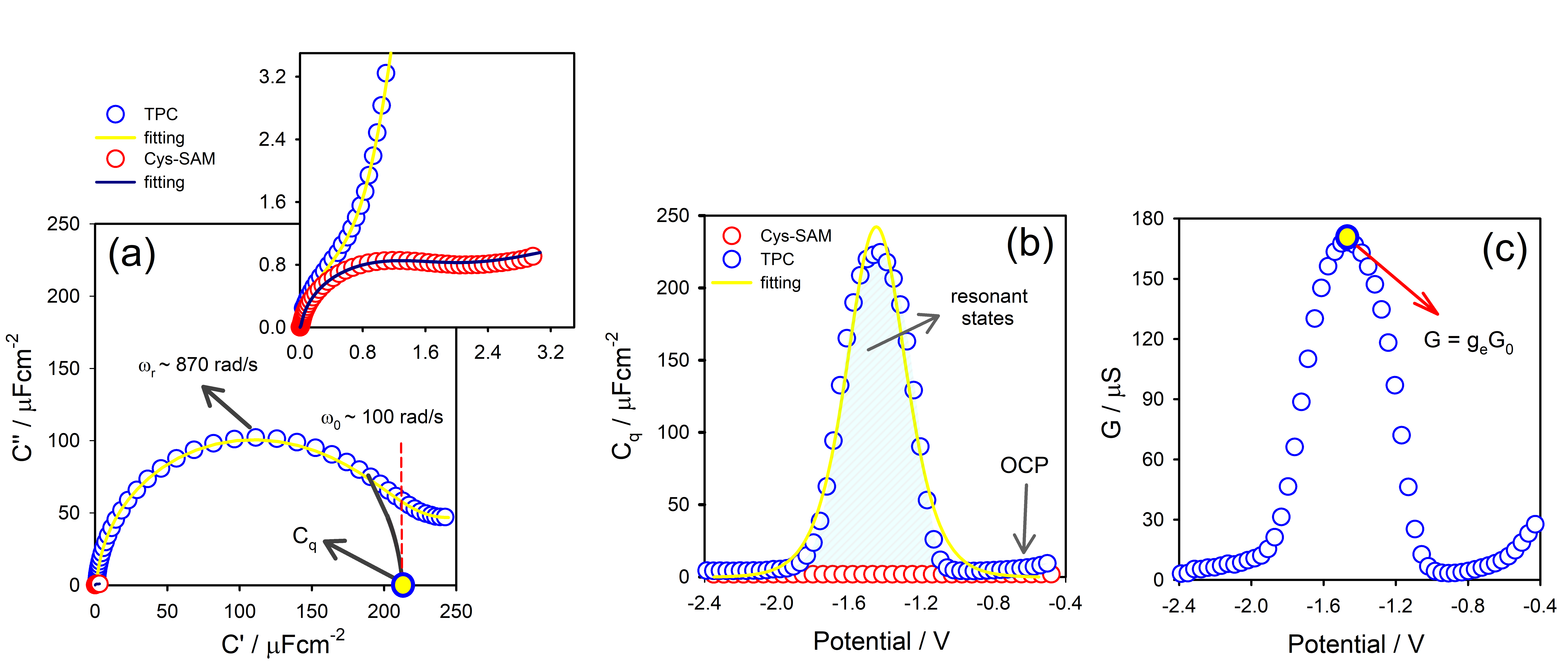}
4\caption{(a) Nyquist capacitive diagrams were recorded at the Fermi level of the electrode, denoted as $E_F/e$, corresponding to a potential bias of $-$1.43 V \textit{versus} a platinum pseudo-reference. The electrode was composed of either cysteamine (Cys SAM) alone or containing \textbf{TPC}-Au on top. Electrochemical measurements were conducted in a 20 mmol L$^{-1}$ concentration of TBAClO$_4$ in DMF. The $C_q$ value of junctions containing \textbf{TPC} (measured at $\omega_0 \sim$ 100 rad/s) is two orders of magnitude higher than the capacitance of the junction solely comprising cysteamine. (b) By measuring the capacitance of junctions containing only cysteamine or with \textbf{TPC} on top, it is evident that the quantum relaxation dynamics and the relativistic communication of states between the electrode and the thiophene group contained in the junction containing \textbf{TPC} give rise to a $C_q$ state associated with a conductance quantum of $g_s e^2/h$. (c) The response of quantum conductance $G \sim \omega_0 C''$ of the junction containing \textbf{TPC} is depicted as a function of the energy of the electrode, where $C''$ represents the imaginary component of the capacitance of the interface (Eq.\ref{eq:Complex-Cq}). As anticipated from the analysis of Eq.\ref{eq:Cq-thermal} and Eq.~\ref{eq:k-finiteT}, $G$ exhibits a Gaussian-shaped behavior. Furthermore, at the Fermi level, both $C_q$ and $G$ reach their maximum values. The maximum value of $G$ attained at the Fermi level was calculated as $g_e G_0$, confirming not only the degeneracy of energy $g_e$ as predicted by the quantum rate model but also demonstrating the adiabatic establishment of relativistic quantum electrodynamics for the dynamics of the \textbf{TPC} junction studied in this work.}
\label{fig:EIS+DOS-FLevel+QC}
\end{figure*}

Note also that the equilibrium charge relaxation frequencies of the cysteamine and the \textbf{TPC} molecular junction are very different. The DOS of the cysteamine was resolved as at $\omega_0 \sim$ 565 rad/s, whereas for the \textbf{TPC} junction, it was at $\omega_0 \sim$ 60 rad/s. The Gaussian-shaped response of the \textbf{TPC} molecular junction confirms the theory not only because it indicates the existence of the displacement electric current associated with interfacial quantum states but because these are electronic quantum mechanical states measured at room temperature. There is a good agreement ($R^2 \sim$ 0.998) with the quantum rate theory, as verified in Figure~\ref{fig:EIS+DOS-FLevel+QC}\textit{b}, where the fitting (yellow line) of Eq.~\ref{eq:Cq-thermal} to the experimental thermal broadened DOS curve (blue circles) is demonstrated. It is worth noting that at the Fermi-level ($E_F$ = -1.43 V) of the junction, corresponding to an occupancy of $f = 1/2$ in Eq.~\ref{eq:Cq-thermal}, $C_q$ is resolved as $C_q = e^2 N/4k_B T$. Owing to $C_q$, at Fermi-level, is experimentally measured as the maximum of the Gaussian-like response of Figure~\ref{fig:EIS+DOS-FLevel+QC}\textit{b}, as predicted by Eq.~\ref{eq:Cq-thermal}, the total number of quantum states $N$ is thus calculated as $N = 4 k_B T C_q/e^2$ $\sim$ \num{6.3e12}.

This value of $N \sim \num{6.3e12}$, calculated at the Fermi level, corresponds to a quarter of the total number of states owing to that the expression $N = 4 k_B T C_q/e^2$ does not take into account the energy $E = e^2/C_q$ degeneracy which, in agreement to the quantum rate theory, must be taken into account for an appropriate interpretation and quantification of the quantum electrodynamics of communication with the electrode existing via the ambipolar nature of the electric displacement current. Hence, by calculating the total amount of states as $N = 1/e\int C_q(V) dV$, which is a pure mathematical analysis without a physical interpretation where the integral $\int_{0}^{V} C_q(V)$ corresponds to the area under the curve of the Gaussian-shaped response of $C_q$ (Figure~\ref{fig:EIS+DOS-FLevel+QC}\textit{b}), a value of $\sim$ \num{25.4e12} is obtained, which is, as expected, approximately four times higher than the value calculated by $N = 4 k_B T C_q/e^2$. The above-stated analysis confirms the physical interpretation of the energy degeneracy and electric current displacement required to interpret the physical chemistry of the electronic communication of the \textbf{TPC} molecular states with the electrode. Finally, note that a consideration of the energy degeneracy would theoretically lead to $N = k_B T C_q/e^2$ at the Fermi level, which fits quite well to the mathematical analysis of the DOS curve of Figure~\ref{fig:EIS+DOS-FLevel+QC}\textit{b}. 

To understand that the electrode can act as both $D$ or $A$ states for the thiophene group of the \textbf{TPC} organic compounds communicating with the electrode, as illustrated in Figure~\ref{fig:electrode-DA}, note the bias potential applied above or below the Fermi level of the junction. Assuming a negative elementary charge for the electron and the Fermi level as the reference, as depicted in Figure~\ref{fig:electrode-DA}, a positive potential bias of the electrode acts as $D$ states for the \textbf{TPC} molecular states and a negative bias as $A$. Note also that the displacement electric current can only be measured, as the case of redox junctions, by using a time-dependent perturbation. 

The above analysis demonstrates that the electron transport (or wave communication phenomenon) is intrinsically associated to a quantum electrodynamic event investigated only under an appropriate time-dependent electric potential perturbation. The Gaussian-shaped DOS resolved for an ensemble of \textbf{TPC} organic molecules assembled to the electrode represents a particular quantum mechanical event not yet well-understood. In other words, albeit this electrodynamic phenomenon follows the well-known Einstein-Planck $E = h\nu$ phenomenology, which is quite specific for the physics of high energy particles that precisely quantifies and explains the photon-electric phenomenon, the electronic communication established and measured here between \textbf{TPC} molecular states and electrode is a low energy electrodynamics in which the intrinsic frequency of the quantum RC process corresponds to an ultra-low frequency of the electromagnetic spectrum.

For instance, in the present experimental study, the $\nu$ characteristic frequency for thiophene electronic states to communicate with the electrode corresponds to frequencies $\sim$ 140 s$^{-1}$ or 870 rad s$^{-1}$, as indicated in Figure~\ref{fig:EIS+DOS-FLevel+QC}\textit{a}, meaning an electromagnetic phenomenon with energies $E = h\nu$ in between peV to feV. Note that this characteristic low-energy quantum RC dynamics is here investigated using a small amplitude perturbation of 10 mV RMS, meaning an amplitude perturbation of about seven mV peak-to-peak, corresponding to an electric-field amplitude perturbation that is about three times lower than that of the thermal voltage, i.e. $k_BT/e \sim$ 25.7 mV. In other words, despite a thermal energy ($k_BT$) three times higher than that used for taking information from the experiment, there is a phase resolution of the electric signal. The latter is explained based on the existence of an experimental quantum coherence despite the thermodynamics and of the entropy.

As previously noticed, the communication between \textbf{TPC} molecule moieties and electrode states does not involve the transfer of electrons nor electrochemical reactions limited by diffusion. The latter conjunction of facts effectively demonstrates that the low-frequency electromagnetic perturbation of the electrode permits a resonant communication with thiophene molecular states through an electric displacement current that is essentially dynamic. This electromagnetic communication between electrons perturbed in the electrode and the electric current resonance in the thiophene moiety of the \textbf{TPC} compound is additionally an adiabatic type of electric communication, as was previously observed in measurements conducted in single-layer graphene~\cite{Bueno+Davis-2020}.

This section demonstrated that by applying a time-dependent potential perturbation (with an amplitude that is at least three times lower than the thermal energy), with a specific frequency of $\omega_0$, in the electrode, there is an induced electric and magnetic field that implies a low-energy electromagnetic wave propagation from the electrode toward \textbf{TPC} molecules. This electromagnetic wave can perturb the quantum resonant states of the thiophene in the \textbf{TPC} molecular film, implying a low-energy electrodynamic way of studying electronics. The response to the sinusoidal potential perturbation of the electrode measured as a time-dependent induced electric current (referred as to displacement current) associated with the meaning of $C_q$ obeys the quantum RC dynamics of Eq.~\ref{eq:Complex-Cq}. 

The physical interpretation for this observed low-energy electrodynamic phenomenon is that there is a quantum coherence phase dynamics between the potential perturbation and electric signal responses, not affected by the thermodynamics. Using the $E = h\nu$ electrodynamics interpretation of this electric perturbation or spectroscopic analysis, it implies a viewpoint of the phenomenon governed by the propagation of a $E = h\omega_0$ electromagnetic wave energy (a massless transport of energy that resembles an antenna electromagnetic communication) toward the \textbf{TPC} molecular film that, contrary to the sole cysteamine chemistry of the interface, has intra-molecular resonant dynamics owing to the present thiophene-pyrrole aromatic rings within the film. Quite interesting is that the quantum coherence at room temperature allows for perturbing the quantum states (with extremely low-energy) spectroscopic energy that permits studying of the quantum mechanics of push-pull compounds in its almost pristine electrodynamics (there is no perturbation or interaction of a high energy particle with the electrons of the system, as is the case of traditional electron or photon spectroscopic measurements). The nature of this electromagnetic perturbation allows us to idealize a new electric spectroscopic way of characterizing the electronic structure and to measure the DOS through direct measurement of $C_q$, as was predicted in reference~\citep{Miranda-2016}.

This spectroscopic methodology was referred to as quantum-rate spectroscopy (QRS) and allowed us to measure the electronic structure of graphene~\cite{Lopes-2023} and inorganic quantum dots~\cite{pinzon2023quantum} at room temperature and electrolyte medium with an accuracy that is comparable to those obtained by angle-resolved photoemission (ARPES) and scanning tunneling (STS) spectroscopic methods that require low-temperature and ultra-high vacuum environment besides of expensive piece of equipment whereas QRS use hand-held inexpensive potentiostats. Furthermore, calculated by density-functional computational methods (DFT) agreed with QRS spectra. DFT computational methods were employed to model the electronic structure of molecular junctions, where the equilibrium charging dynamics of redox molecular junctions were studied~\cite{Feliciano-2020, bueno2021density, bueno2015capacitance}. Conceptual DFT can be easily correlated with the meaning of $C_q$ as the energy $e^2/C_q$ corresponds to the variation of the chemical potential upon the exchange of electronic particles~\cite{Miranda-2019}. Furthermore, $C_q$ can serve as an experimental functional of the electron density~\cite{Bueno-2017-2} and a detailed theoretical approach for the meaning of electrochemical capacitance\footnote{The meaning of the electrochemical capacitance incorporates other contributions besides the energy of the quantum states, such as electrostatics. In the case of Tobias's work~\citep{Tobias-2023}, it also incorporates exchange-correlation contributions as a term within DFT's Hamiltonian.} based on conceptual DFT was also recently demonstrated by others~\cite{Tobias-2023}. The correlation between $C_q$ and electron transport properties was previously established within a DFT approach~\cite{Bueno-2017}. However, to better understand the electrodynamics within the quantum-rate viewpoint, time-dependent DFT is probably the right approach for future works.

In the next section, it will be discussed, in more detail, the adiabatic nature of the above-discussed quantum electromagnetic phenomenon.

\subsection{Adiabatic Quantum Electrodynamics of Push-Pull Molecular Junctions}\label{sec:Adiab-QEDs}

An analysis of the adiabatic nature of the quantum electromagnetic phenomenon discussed in section~\ref{sec:QED-roomT} enables investigating the efficiency of the electrodynamics of communication of measured DOS of thiophene-pyrrole aromatic rings with the electrode. The confirmation that there is an adiabatic electron coupling signifies that there is a maximum efficiency of electron `transport' or better said, transmittance. Effectively, as demonstrated in preceding sections, there is no mass diffusional transport of species and the phenomenon follows a relativistic Dirac-like electrodynamics~\citep{Dirac-1928}, in agreement to Eq.~\ref{eq:Planck-Einstein} and experiments conducted in graphene~\citep{Bueno-2022}, implying that for push-pull molecular junctions there is a relativistic quantum electrodynamics equivalent to dynamics followed by massless Fermionic particle within a quantum rate theoretical interpretation of the quantum electrodynamics of graphene. 

In other words, within the quantum rate analysis of Eq.~\ref{eq:Planck-Einstein}, there is the meaning of $\nu = e^2/hC_q = E/h$, as stated in Eq.~\ref{eq:nu}, that, with the consideration of $g_e$ electric-field or potential degeneracy owing to an electron-ion pair dynamics within the electrolyte, leads to define a quantum electrodynamics frequency that follows Eq.~\ref{eq:nu-degenerated}. This electrodynamics frequency, according to Eq.~\ref{eq:nu-degenerated}, is proportional to the resistance quantum $R_q = 1/G_0 = h/2e^2 \sim$ 12.9 k$\Omega$ of the molecular junction.

\begin{figure}[h]
\centering
\includegraphics[height=5.5cm]{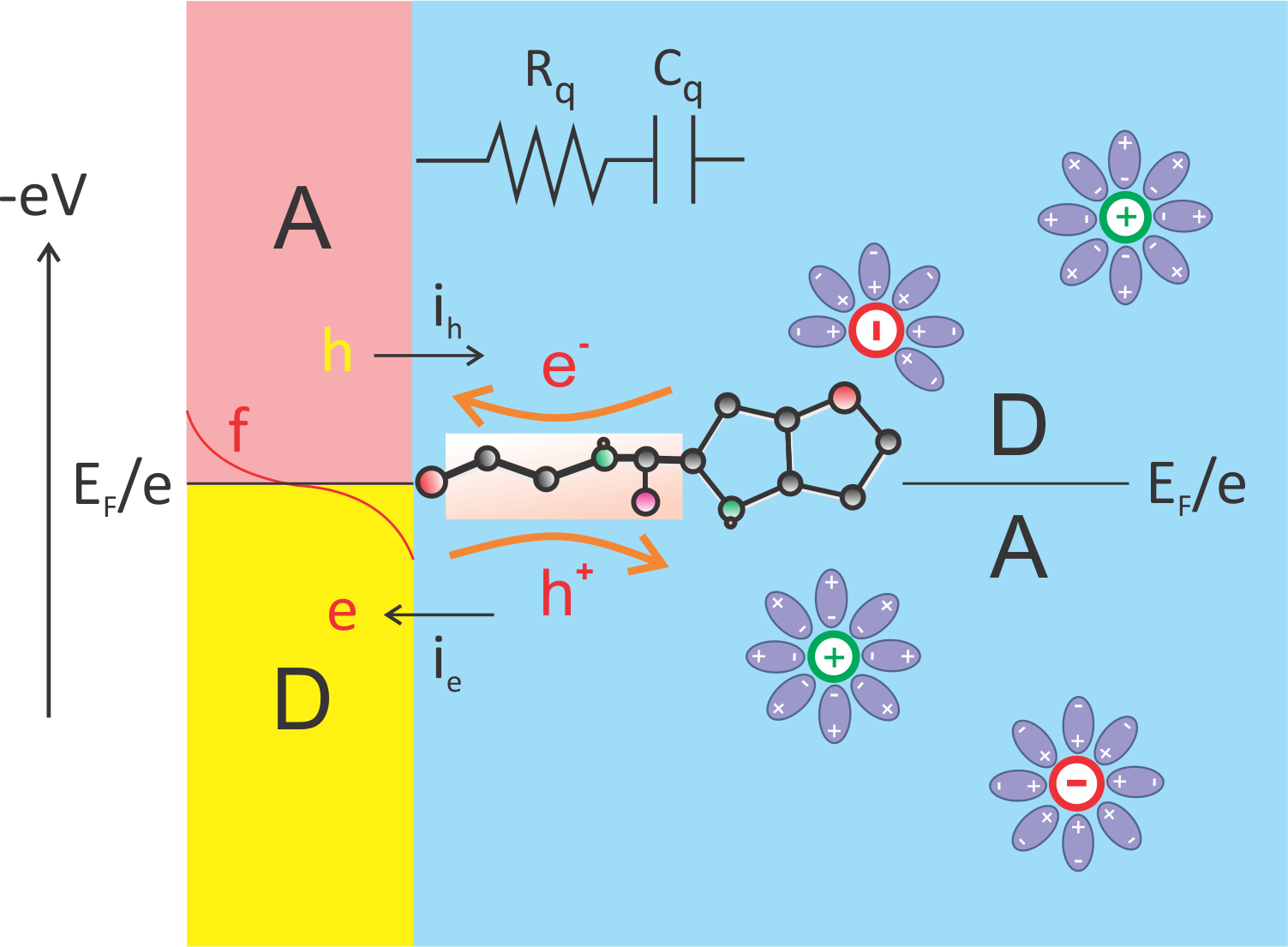}
\caption{The electron/hole communication dynamics of the electrode with the thiophene group of \textbf{TPC} molecular film depends not only on the existence of a potential perturbation in the electrode but also on the bias for the Fermi potential $E_F/e$ of the junction. Depending on the Fermi energy $E_F$ level of the electrode, there is an electron or hole current, and the thiophene group of the (\textbf{TPC} can act as a $D$ or $A$ level, for instance. For positive bias of potential (providing the elementary charge of the electron is assumed as negative value) concerning $E_F/e$, the thiophene groups of \textbf{TPC} molecular film act as $D$ and the electrode as $A$ state and \textit{vice-versa}. In this schematic depicting of the quantum electrodynamics, there is an electric hole current from the electrode to thiophene, and `effectively' an electronic communication of the electrode with the molecular film, as illustrated in this figure.}
\label{fig:electrode-DA}
\end{figure}

This resistance was here shown to be adiabatic, implying a perfect transmittance of electrons through a resistance quantum $R_q$ of 12.9 k$\Omega$ within the junction without any potential decaying term, meaning that $\kappa = \exp(-\beta L) \sim 1$ and demonstrating a perfect electronic coupling $\kappa$ between molecular $D/A$ quantum and electrode $A/D$ energy states. For redox reactions~\citep{Bueno-2023-3}, as studied within the quantum rate theory, there is a charge transfer resistance~\citep{Sanchez-2022-1} that is equivalent to the quantum resistance of $R_q = N \exp(-\beta L)/G \sim$ 12.9 k$\Omega$ of the electrochemical junction undergoing redox reactions within an electron tunneling dynamics that contains a potential decaying term of $\exp \left( -\beta L \right)$, where $G$ is the total (measurable) conductance of the junction and $N$ the total (measurable) number of quantum states participating in the redox reaction, $L$ is the length of the barrier and $\beta$ (both experimentally estimated) is a parameter associated with the potential barrier height.

The present study of the molecular electronics and electrodynamics of push-pull heterocyclic resonant junctions demonstrates that the $R_q$ measured at the Fermi level ($E_F/e = -$1.43 V) is astonishingly close to the theoretical value of 12.9 k$\Omega$, and contrary to redox-active junctions, is not associated to a tunneling mechanism and hence there is no $\exp (-\beta L)$ decaying term. For the case of push-pull heterocyclic resonant junctions, the quantum conductance $G$ corresponds to $g_eG_0$ (see SI document, section S4 for additional information) and the communication of the quantum (resonant) states of the molecules with the electrode is adiabatic and do not depend on the properties of the linker, which is, in the present junction, a chemical $\pi$-type bridge.

The above statement confirms the findings from the equivalent circuit analysis of \textbf{TPC} molecular junctions. The equivalent circuit was constructed using a circuital mathematical fitting approach for junction modeling. This circuit predicts a dominant quantum RC dynamics over the classical polarization dynamics of the interface, consistent with Eq.\ref{eq:Complex-Cq}. Accordingly, the capacitive complex function depicted in Figure\ref{fig:EIS+DOS-FLevel+QC}\textit{a} was fitted to the equivalent circuit shown in Figure~\ref{fig:circuit-comparisons}\textit{b}. In this model, the polarization dynamics of the interface, including those associated with the cysteamine underlying film (Cys-SAM), were considered to increase the accuracy of the analysis. Mathematically, at the Fermi level of the junction, the RC response is predominantly governed by the quantum RC term.

The fitting (yellow line) of the complex capacitance function (blue dots) of the interface to the circuit in Figure~\ref{fig:circuit-comparisons}\textit{b} is illustrated in Figure~\ref{fig:EIS+DOS-FLevel+QC}\textit{a}, demonstrating a suitable numerical mathematical fitting. A value of $\sim$ 225 $\mu$F cm$^{-2}$ was obtained for $C_q$, while a value of $\sim$ 2.3 $\mu$F cm$^{-2}$ was assigned to the capacitive $C_{nf}$ contribution of the cysteamine film, consistent with the equivalent circuit in Figure~\ref{fig:circuit-comparisons}\textit{b}. The minimal $C_{nf}$ contribution was confirmed by separately fitting (black line) the complex capacitive spectrum of the cysteamine film (red dot), as shown in the inset of Figure~\ref{fig:EIS+DOS-FLevel+QC}\textit{a}, to the equivalent circuit highlighted in the gray box of Figure~\ref{fig:circuit-comparisons}\textit{b}. The average values obtained for each circuital element of the equivalent circuit analysis are presented in Table~\ref{table:example-1}. Therefore, this  equivalent circuit analysis demonstrates that the quantum RC dynamics govern over the non-faradaic polarization charging dynamics at the Fermi level.

It is noteworthy that the resistance quantum of molecular junctions embedded in an electrolyte environment, as observed in the case of redox-active molecular junctions~\citep{Sanchez-2022-1, Bueno-2023-3} undergoing electrochemical reaction, comprises the total series resistance of the interface and not only the resistance of the \textbf{TPC}-cysteamine film. Specifically, $R_q = R_s + R_{qt}$, where $R_s$ includes the contact and solution resistance, and $R_{qt}$ is an internal resistance within the molecular film, sometimes referred to as uncompensated resistance~\citep{Bueno-2013}. In other words, the resistance quantum of the interface encompasses the total series resistance of the junction and not just the resistance of the \textbf{TPC}-cysteamine film assembled over the electrode.

\begin{table}[h]
\small
\caption{Values of circuit parameters were derived from the fitting of the impedance spectra to the equivalent circuit illustrated in Figure~\ref{fig:Lmode+circuit}\textit{b}. Here, $\overline{x}$ denotes the mean average value, and $\sigma_{\overline{x}}$ represents the standard error of the mean of three independently constructed junctions incorporating \textbf{TPC} molecules anchored to a cysteamine film.
}
  \label{table:example-1}
  \begin{tabular*}{0.48\textwidth}{@{\extracolsep{\fill}}lll}
    \hline
    Parameter & $\overline{x}$ & $\sigma_{\overline{x}}$ \\
    \hline
    $R_s$(k$\Omega$) & 1.13& 0.02 \\
    $C_{m}$ ($\mu$F cm$^{-2}$) & 0.056 & 0.008 \\
    $R_{nf}$ (k$\Omega$) & 3.0 x 10$^{-4}$ & 0.5 x 10$^{-4}$ \\
    $C_{nf}$ ($\mu$F cm$^{-2}$) &    2.3 & 0.3 \\
    $R_{qt}$ (k$\Omega$) & 0.93 & 0.05 \\
    $C_q$ ($\mu$F cm$^{-2}$) & 224 & 16 \\
    \hline
  \end{tabular*}
\end{table}

Note also that the characteristic angular frequency associated to $\nu$ is ascribed to as $\omega = e^2/\hbar C_q$ which, in agreement to $\omega \tau$ nomenclature stated in Eq.~\ref{eq:Complex-Cq}, leads to $\tau = 2\pi R_qC_q$, where $R_q = R_s + R_{qt}$, such as that $R_q = 2\pi(R_s + R_{qt})$ \cite{sanchez2022quantum}. The values of $R_s$ and $R_{qt}$ computed from the fitting to the equivalent circuit, as shown in Table~\ref{table:example-1}, conduct to $R_q \sim$ 12.92 $\pm$ 0.44 k$\Omega$. This experimental value of $R_q$, within the experimental error, is indubitably very close to the theoretical value of 12.99 k$\Omega$ and demonstrates the certainty of the quantum rate theory and the adiabatic character of the quantum electrodynamics of push-pull molecular junctions.

In other words, the above analysis demonstrates that the measured resistance at the Fermi level of the \textbf{TPC}-electrode junction, embedded in an electrolyte medium, obeys the quantum limit of $R_q = h/2e^2 \sim$ 12.99 k$\Omega$ and confirms that there is a maximum quantum mechanical efficiency of the electron transport in this type of organic molecular junctions with an adiabatic electron coupling regime that follows relativistic dynamics, fulfilling Eq.~\ref{eq:Planck-Einstein}. The analysis do not only validates the quantum rate analysis of these types of junctions but also predicts a Fermionic massless particle dynamics that conforms to the electrodynamics of graphene, as previously studied using a quantum rate theoretical approach.

Besides the calculation of $R_q \sim$ 12.92 k$\Omega$ via the resistive elements that comprise the \textbf{TPC} molecular junction using an equivalent circuit analysis, as conducted above, the value of $R_q = 1/G_0$ is alternatively quantified through an analysis of $\nu = g_e g_s e^2/hC_q = g_e (G_0/C_q)$ measured at the Fermi level of the junction, from where the measurements of $\nu$ and $C_q$ permits to evaluate $G_0$, self-consistently demonstrating that $R_q$ is $R_q = g_e / \nu C_q$. Average values of $\nu = \omega/2\pi$ and $C_q$, at the Fermi level, were measured in three independently fabricated junctions, as detailed in section S3 of the SI document. $C_q$, obtained at the maximum of the Gaussian-shaped response, corresponded to the value of $C_q$ at the peak of the DOS function (measured at a frequency of $\omega_0$), as shown in Figure~\ref{fig:EIS+DOS-FLevel+QC}\textit{b}. The average value of $G_0$ estimated from the above analysis was 76 $\pm$ 3 $\mu$S that, within experimental error, is close to the theoretical value of $G_0 =$ 77.5 $\mu$S and confirms the certain of the analysis.

Still, another analysis reveals the certainty of the quantum rate theory, consisting in calculating the quantum of the conductance $G$ of the junction directly from the complex capacitance associated with Eq.~\ref{eq:Complex-Cq}, from where $G = \omega_0 C''$ is obtainable. The measurement of $G$ at $\omega_0$, as a function of electrode potential level, is shown in Figure~\ref{fig:EIS+DOS-FLevel+QC}\textit{c}. Such as in the case of the $C_q$, shown in Figure~\ref{fig:EIS+DOS-FLevel+QC}\textit{b}, $G$ has a maximum value at the Fermi level of the junction. The peak of $G$ at the Fermi level has a value of $\sim$ 154.7 $\pm$ 2.0 $\mu$S, corresponding to a mean value (among three independently fabricated junctions) that is quite close to the theoretical value of $g_e G_0 \sim$ 154.9 $\mu$S, i.e. twice $G_0 \sim$ 77.5 $\mu$S. Precisely, as shown in more detail in the S4 section of the SI document, the mean value of $G$ obtained from three different junctions corresponds to 154.7 $\pm$ 2.0 $\mu$S, corresponding to about 1\% of difference with the theoretical $g_e G_0 \sim$ 154.9 $\mu$S value.

The analysis conducted in the last paragraph for the calculation of $G$ as $G = \omega_0 C''$ not only demonstrated the confidence of a quantum rate theoretical analysis of electrodynamics of push-pull organic molecular junctions but also confirmed that the $g_e$ is critical for the analysis of molecular junctions embedded in an electrolyte environment.

Finally, the consistency is confirmed of a comparative analysis of irreversible and reversible electric modes of push-pull moiety electronic states to communicate with the electrode, which can be conducted using different electrochemical settings, as discussed in section~\ref{sec:Adiab-QEDs}. Particularly in the schemes illustrated in Figure~\ref{fig:EC-settings}, there is an irreversible (scheme depicted in Figure~\ref{fig:EC-settings}\textit{a}) or reversible (in Figure~\ref{fig:EC-settings}\textit{b}) way of electronic communication of the push-pull molecular states with the electrode. 

In the irreversible (kinetically controlled by diffusion) scheme, in which the molecules are freely in an electrolyte environment, there is an electric potential separation of $\sim$ 2.08 V between the first oxidation (electrons flowing from the HOMO state of the molecules to the electrode) and first reduction (electrons flowing from the electrode to the LUMO state of the molecules) current peaks, which have not an ambipolar character but can be associated with $\Delta E_{H-L}$ gap, as introduced in section~\ref{sec:introduction}. On the other hand, in the case of a reversible resonant communication of the molecular states with the electrode, which occurs thankfully due to a chemical attachment of the molecules to the electrode (with diffusionless kinetics), the electric current is intrinsically ambipolar, as illustrated in Figure~\ref{fig:electrode-DA}. In this case, there is a resonant adiabatic quantum electrodynamics, as concluded in the present section, implying, as was demonstrated in a previous work~\citep{Bueno-book-2018}, that the energy $E = e^2/C_q$ is related to the HOMO and LUMO gap of molecular ensemble attached to the electrode. This $\Delta E_{H-L}$ gap is estimated as the width of the DOS depicted in Figure~\ref{fig:EIS+DOS-FLevel+QC}\textit{b}, hence computed as $\sim$ 1 eV, corresponding approximately to the half of $\Delta E_{H-L}$ gap of $\sim$ 2.08 eV estimated from the CV pattern measured in an irreversible electrochemical setting. 

The above analysis strengthens the interpretation of the degeneracy $g_e$ as being due to the presence of an ambipolar electric current associated with the overlap of quantum ($C_q$) and classical ($C_e$) capacitive states of the junction. Equivalently, this can be interpreted as an energy degeneracy linked to the electric potential ($e/C_e \sim e/C_q$) arising from the electric-field screening conducted by the electrolyte medium over the quantum resonant states of the push-pull moieties communicating with the electrode.

\section{Final Remarks and Conclusions}
\label{sec:conclusions}

The present study demonstrates the feasibility of establishing an adiabatic electron coupling between a push-pull molecular ensemble and a probe electrode. This adiabatic electron coupling enables the investigation of the quantum resonant states of the molecules with the electrode. The exploration of quantum states resonating with the electrode occurs through an electromagnetic communication, adhering to the premises of quantum electrodynamics, which follow the Einstein-Planck relationship $E = h\nu$. This implies a quantum relativistic dynamics between quantum states in push-pull heterocyclic sites and the electrode, wherein, according to the quantum rate theory, the $\nu = e^2/hC_q$ rate is directly linked to the quantum capacitance $C_q \propto e^2$DOS of the junction. This capacitance is proportional to the accessible density of junction quantum resonant states, where DOS $= (dn/dE)$.
 
As $C_q$ is experimentally accessible through conventional impedance spectroscopic measurements such as impedance or impedance-derived capacitance spectra (within the interpretation of Eq.~\ref{eq:Complex-Cq}), it provides a way of studying the quantum electrodynamics of molecular junctions embedded in an electrolyte environment. The presence of an electrolyte environment implies an additional electric-field degeneracy $g_e$ that summed to the electron spin degeneracy, conducts to an electromagnetic communication way, demonstrated to follow $E = g_s g_e h\nu$ quantum relativistic electrodynamics, permitting demonstration that $\nu = g_e G/C_q$, which is not only in agreement to the quantum rate theory but also establish an alternative way of designing molecular electronics junctions and nanoscale electronic circuits.
 
The observed quantum relativistic phenomenology of \textbf{TPC} push-pull molecular junctions is equivalent to that observed for electro-active interfaces undergoing redox reactions; however, for the case of \textbf{TPC} junctions there is a particular adiabatic electron coupling in which effectively $\nu = g_e G_0/C_q$ is obeyed, instead of $\nu = g_e G/C_q$, as observed for redox-active molecular junctions, where a $G \propto \exp (-\beta L)$ exists. The existence of $G \propto \exp (-\beta L)$ for redox-active junctions, thus with a tunneling mechanism governing the electronic coupling, is different from that observed in push-pull that implies that $G = G_0$, possibly associated with the $\pi$-type molecular bridge electronic coupling. In one (electrochemical) or another (electronic) situation, the coupling of states to the electrode obeys a relationship consistent with the quantum rate $\nu = g_e G/C_q$ concept.
 
\section*{Author Contributions}
\textbf{Edgar Fabian Pinzón Nieto}: Methodology, Investigation, Visualization, Writing - Original Draft. \textbf{Laís Cristine Lopes}: Methodology, Investigation, Visualization. \textbf{Adriano dos Santos}: Methodology and Investigation. \textbf{Maria Manuela Raposo}: Methodology, Investigation, Visualization, Synthesis of \textbf{TPC} Compound, Writting original-draft of the manuscript. \textbf{Paulo Roberto Bueno}: Conceptualization, Theoretical Analysis, Supervision, Resources, Methodology, Writing - Original Draft and Final Versions of this text. 

\section*{Conflicts of interest}
There are no conflicts to declare

\section*{Acknowledgements} 
Paulo R. Bueno: Grateful to the Sao Paulo State Research Foundation (FAPESP) for grants 2017/24839-0 and the National Council for Scientific and Technological (CNPq).

Edgar F. Pinzón: Acknowledges 
FAPESP for his PhD scholarship (grant 2018/24525-9).

Fundação para a Ciência e Tecnologia (FCT) and FEDER-COMPETE: For financial support through the research center CQUM (UID/QUI/0686/2016) and (UID/QUI/0686/2020).

National NMR Network (PTNMR): Partially supported by Infrastructure Project Nº 022161, co-financed by FEDER through COMPETE 2020, POCI and PORL, and FCT through PIDDAC.

Microfabrication procedures: Were performed with the assistance of Dr. R. C. Teixeira and the Electronic Packaging Staff at CTI Renato Archer Brazil under proposal PAP : 2019-06568-5.







\printcredits

\bibliographystyle{cas-model2-names}

\bibliography{biblio}

\begin{thebibliography}{53}
\expandafter\ifx\csname natexlab\endcsname\relax\def\natexlab#1{#1}\fi
\providecommand{\url}[1]{\texttt{#1}}
\providecommand{\href}[2]{#2}
\providecommand{\path}[1]{#1}
\providecommand{\DOIprefix}{doi:}
\providecommand{\ArXivprefix}{arXiv:}
\providecommand{\URLprefix}{URL: }
\providecommand{\Pubmedprefix}{pmid:}
\providecommand{\doi}[1]{\href{http://dx.doi.org/#1}{\path{#1}}}
\providecommand{\Pubmed}[1]{\href{pmid:#1}{\path{#1}}}
\providecommand{\bibinfo}[2]{#2}
\ifx\xfnm\relax \def\xfnm[#1]{\unskip,\space#1}\fi
\bibitem[{Alarc{\'o}n et~al.(2021)Alarc{\'o}n, Santos and
  Bueno}]{alarcon2021perspective}
\bibinfo{author}{Alarc{\'o}n, E.V.G.}, \bibinfo{author}{Santos, A.},
  \bibinfo{author}{Bueno, P.R.}, \bibinfo{year}{2021}.
\newblock \bibinfo{title}{Perspective on quantum electrochemistry. a simple
  method for measuring the electron transfer rate constant}.
\newblock \bibinfo{journal}{Electrochimica Acta} \bibinfo{volume}{398},
  \bibinfo{pages}{139219}.
\bibitem[{Bisquert et~al.(2008)Bisquert, Fabregat-Santiago, Mora-Ser{\'o},
  Garcia-Belmonte, Barea and Palomares}]{bisquert2008review}
\bibinfo{author}{Bisquert, J.}, \bibinfo{author}{Fabregat-Santiago, F.},
  \bibinfo{author}{Mora-Ser{\'o}, I.}, \bibinfo{author}{Garcia-Belmonte, G.},
  \bibinfo{author}{Barea, E.M.}, \bibinfo{author}{Palomares, E.},
  \bibinfo{year}{2008}.
\newblock \bibinfo{title}{A review of recent results on electrochemical
  determination of the density of electronic states of nanostructured
  metal-oxide semiconductors and organic hole conductors}.
\newblock \bibinfo{journal}{Inorganica Chimica Acta} \bibinfo{volume}{361},
  \bibinfo{pages}{684--698}.
\bibitem[{Bueno(2018a)}]{bueno2018common}
\bibinfo{author}{Bueno, P.R.}, \bibinfo{year}{2018}a.
\newblock \bibinfo{title}{Common principles of molecular electronics and
  nanoscale electrochemistry}.
\bibitem[{Bueno(2018b)}]{Bueno-book-2018}
\bibinfo{author}{Bueno, P.R.}, \bibinfo{year}{2018}b.
\newblock \bibinfo{title}{The Nanoscale Electrochemistry of Molecular
  Contacts}.
\newblock SpringerBriefs in Applied Sciences and Technology,
  \bibinfo{publisher}{Springer}.
\bibitem[{Bueno(2020)}]{Bueno-2020}
\bibinfo{author}{Bueno, P.R.}, \bibinfo{year}{2020}.
\newblock \bibinfo{title}{Electron transfer and conductance quantum}.
\newblock \bibinfo{journal}{Physical Chemistry Chemical Physics}
  \bibinfo{volume}{22}, \bibinfo{pages}{26109--26112}.
\newblock \DOIprefix\doi{10.1039/d0cp04522e}.
\bibitem[{Bueno(2023a)}]{Bueno-2023-1}
\bibinfo{author}{Bueno, P.R.}, \bibinfo{year}{2023}a.
\newblock \bibinfo{title}{On the electromagnetic nature of planck constant}.
\newblock \bibinfo{journal}{Preprint arXiv:2302.10797} ,
  \bibinfo{pages}{1--5}\DOIprefix\doi{10.48550/arXiv.2302.10797}.
\bibitem[{Bueno(2023b)}]{Bueno-2023-2}
\bibinfo{author}{Bueno, P.R.}, \bibinfo{year}{2023}b.
\newblock \bibinfo{title}{Quantum electromagnetic rate theory of the electron
  and the meaning of the fine-structure constant}.
\newblock \bibinfo{journal}{Preprint arXiv:2302.12886} ,
  \bibinfo{pages}{1--9}\DOIprefix\doi{10.48550/arXiv.2302.12886}.
\bibitem[{Bueno(2023c)}]{Bueno-2023-3}
\bibinfo{author}{Bueno, P.R.}, \bibinfo{year}{2023}c.
\newblock \bibinfo{title}{Quantum rate theory and electron-transfer dynamics: A
  theoretical and experimental approach for quantum electrochemistry}.
\newblock \bibinfo{journal}{Electrochimica Acta} \bibinfo{volume}{466},
  \bibinfo{pages}{142950}.
\newblock \DOIprefix\doi{doi.org/10.1016/j.electacta.2023.142950}.
\bibitem[{Bueno(2024)}]{Bueno-2024}
\bibinfo{author}{Bueno, P.R.}, \bibinfo{year}{2024}.
\newblock \bibinfo{title}{On the fundamentals of quantum rate theory and the
  long-range electron transport in respiratory chains}.
\newblock \bibinfo{journal}{Chemical Society Review}
  \DOIprefix\doi{doi.org/10.1039/D3CS00662J}.
\bibitem[{Bueno et~al.(2021)Bueno, Cruzeiro, Roitberg and
  Feliciano}]{bueno2021density}
\bibinfo{author}{Bueno, P.R.}, \bibinfo{author}{Cruzeiro, V.W.D.},
  \bibinfo{author}{Roitberg, A.E.}, \bibinfo{author}{Feliciano, G.T.},
  \bibinfo{year}{2021}.
\newblock \bibinfo{title}{The density-of-states and equilibrium charge dynamics
  of redox-active switches}.
\newblock \bibinfo{journal}{Electrochimica Acta} \bibinfo{volume}{387},
  \bibinfo{pages}{138410}.
\bibitem[{Bueno and Davis(2020)}]{Bueno+Davis-2020}
\bibinfo{author}{Bueno, P.R.}, \bibinfo{author}{Davis, J.J.},
  \bibinfo{year}{2020}.
\newblock \bibinfo{title}{Charge transport and energy storage at the molecular
  scale: from nanoelectronics to electrochemical sensing}.
\newblock \bibinfo{journal}{Chemical Society Reviews} \bibinfo{volume}{49},
  \bibinfo{pages}{7505--7515}.
\newblock \DOIprefix\doi{10.1039/c9cs00213h}.
\bibitem[{Bueno et~al.(2013)Bueno, Fabregat-Santiago and Davis}]{Bueno-2013}
\bibinfo{author}{Bueno, P.R.}, \bibinfo{author}{Fabregat-Santiago, F.},
  \bibinfo{author}{Davis, J.J.}, \bibinfo{year}{2013}.
\newblock \bibinfo{title}{Elucidating capacitance and resistance terms in
  confined electroactive molecular layers}.
\newblock \bibinfo{journal}{Analytical Chemistry} \bibinfo{volume}{85},
  \bibinfo{pages}{411--417}.
\newblock \DOIprefix\doi{10.1021/ac303018d}.
\bibitem[{Bueno et~al.(2015a)Bueno, Feliciano and Davis}]{bueno2015capacitance}
\bibinfo{author}{Bueno, P.R.}, \bibinfo{author}{Feliciano, G.T.},
  \bibinfo{author}{Davis, J.J.}, \bibinfo{year}{2015}a.
\newblock \bibinfo{title}{Capacitance spectroscopy and density functional
  theory}.
\newblock \bibinfo{journal}{Physical Chemistry Chemical Physics}
  \bibinfo{volume}{17}, \bibinfo{pages}{9375--9382}.
\bibitem[{Bueno and Mercado(2022)}]{Bueno-2022}
\bibinfo{author}{Bueno, P.R.}, \bibinfo{author}{Mercado, D.A.M.},
  \bibinfo{year}{2022}.
\newblock \bibinfo{title}{Quantum rate theory for graphene}.
\newblock \bibinfo{journal}{Journal of Physical Chemistry C}
  \bibinfo{volume}{126}, \bibinfo{pages}{15374--15385}.
\newblock \DOIprefix\doi{10.1021/acs.jpcc.2c02419}.
\bibitem[{Bueno and Miranda(2017a)}]{Bueno-2017-2}
\bibinfo{author}{Bueno, P.R.}, \bibinfo{author}{Miranda, D.A.},
  \bibinfo{year}{2017}a.
\newblock \bibinfo{title}{Conceptual density functional theory for electron
  transfer and transport in mesoscopic systems}.
\newblock \bibinfo{journal}{Physical Chemistry Chemical Physics}
  \bibinfo{volume}{19}, \bibinfo{pages}{6184--6195}.
\newblock \DOIprefix\doi{10.1039/c6cp02504h}.
\bibitem[{Bueno and Miranda(2017b)}]{Bueno-2017}
\bibinfo{author}{Bueno, P.R.}, \bibinfo{author}{Miranda, D.A.},
  \bibinfo{year}{2017}b.
\newblock \bibinfo{title}{Conceptual density functional theory for electron
  transfer and transport in mesoscopic systems}.
\newblock \bibinfo{journal}{Physical Chemistry and Chemical Physics}
  \bibinfo{volume}{19}, \bibinfo{pages}{6184--6195}.
\newblock \DOIprefix\doi{10.1039/C6CP02504H}.
\bibitem[{Bueno et~al.(2015b)Bueno, Schrott, Bonanni, Simison and
  Busalmen}]{Bueno-2015}
\bibinfo{author}{Bueno, P.R.}, \bibinfo{author}{Schrott, G.D.},
  \bibinfo{author}{Bonanni, P.S.}, \bibinfo{author}{Simison, S.N.},
  \bibinfo{author}{Busalmen, J.P.}, \bibinfo{year}{2015}b.
\newblock \bibinfo{title}{Biochemical capacitance of geobacter sulfurreducens
  biofilms}.
\newblock \bibinfo{journal}{ChemSusChem} \bibinfo{volume}{8},
  \bibinfo{pages}{2492--2495}.
\bibitem[{Bure{\v{s}}(2014)}]{Burevs-2014}
\bibinfo{author}{Bure{\v{s}}, F.}, \bibinfo{year}{2014}.
\newblock \bibinfo{title}{Fundamental aspects of property tuning in push--pull
  molecules}.
\newblock \bibinfo{journal}{RSC Advances} \bibinfo{volume}{4},
  \bibinfo{pages}{58826--58851}.
\bibitem[{B{\"u}ttiker et~al.(1993)B{\"u}ttiker, Thomas and
  Pr{\^e}tre}]{buttiker1993mesoscopic}
\bibinfo{author}{B{\"u}ttiker, M.}, \bibinfo{author}{Thomas, H.},
  \bibinfo{author}{Pr{\^e}tre, A.}, \bibinfo{year}{1993}.
\newblock \bibinfo{title}{Mesoscopic capacitors}.
\newblock \bibinfo{journal}{Physics Letters A} \bibinfo{volume}{180},
  \bibinfo{pages}{364--369}.
\bibitem[{Cai et~al.(2013)Cai, Guo, Yang, Peng, Luo, Liu, Zhang, Liu and
  Zhang}]{Cai-2013}
\bibinfo{author}{Cai, Z.}, \bibinfo{author}{Guo, Y.}, \bibinfo{author}{Yang,
  S.}, \bibinfo{author}{Peng, Q.}, \bibinfo{author}{Luo, H.},
  \bibinfo{author}{Liu, Z.}, \bibinfo{author}{Zhang, G.}, \bibinfo{author}{Liu,
  Y.}, \bibinfo{author}{Zhang, D.}, \bibinfo{year}{2013}.
\newblock \bibinfo{title}{New donor--acceptor--donor molecules with pechmann
  dye as the core moiety for solution-processed good-performance organic
  field-effect transistors}.
\newblock \bibinfo{journal}{Chemistry of Materials} \bibinfo{volume}{25},
  \bibinfo{pages}{471--478}.
\bibitem[{Cardona et~al.(2011)Cardona, Li, Kaifer, Stockdale and
  Bazan}]{Cardona-2011}
\bibinfo{author}{Cardona, C.M.}, \bibinfo{author}{Li, W.},
  \bibinfo{author}{Kaifer, A.E.}, \bibinfo{author}{Stockdale, D.},
  \bibinfo{author}{Bazan, G.C.}, \bibinfo{year}{2011}.
\newblock \bibinfo{title}{Electrochemical considerations for determining
  absolute frontier orbital energy levels of conjugated polymers for solar cell
  applications}.
\bibitem[{Castro et~al.(2016)Castro, Belsley and Raposo}]{castro2016synthesis}
\bibinfo{author}{Castro, M.C.R.}, \bibinfo{author}{Belsley, M.},
  \bibinfo{author}{Raposo, M.M.M.}, \bibinfo{year}{2016}.
\newblock \bibinfo{title}{Synthesis and characterization of push--pull
  bithienylpyrrole nlophores with enhanced hyperpolarizabilities}.
\newblock \bibinfo{journal}{Dyes and Pigments} \bibinfo{volume}{131},
  \bibinfo{pages}{333--339}.
\bibitem[{Dirac(1928)}]{Dirac-1928}
\bibinfo{author}{Dirac, P.A.M.}, \bibinfo{year}{1928}.
\newblock \bibinfo{title}{The quantum theory of the electron}.
\newblock \bibinfo{journal}{Proceedings of the Royal Society A: Mathematical,
  Physical and Eng. Sci.} \bibinfo{volume}{117}, \bibinfo{pages}{610–624}.
\bibitem[{Feliciano and Bueno(2020)}]{Feliciano-2020}
\bibinfo{author}{Feliciano, G.T.}, \bibinfo{author}{Bueno, P.R.},
  \bibinfo{year}{2020}.
\newblock \bibinfo{title}{Two-dimensional nature and the meaning of the density
  of states in redox monolayers}.
\newblock \bibinfo{journal}{Journal of Physical Chemistry C}
  \bibinfo{volume}{124}, \bibinfo{pages}{14918--14927}.
\newblock \DOIprefix\doi{10.1021/acs.jpcc.0c04598}.
\bibitem[{Fernandes et~al.(2021)Fernandes, Castro, Ivanou, Mendes and
  Raposo}]{Fernandes-2021}
\bibinfo{author}{Fernandes, S.S.}, \bibinfo{author}{Castro, M.C.R.},
  \bibinfo{author}{Ivanou, D.}, \bibinfo{author}{Mendes, A.},
  \bibinfo{author}{Raposo, M.M.M.}, \bibinfo{year}{2021}.
\newblock \bibinfo{title}{Push-pull heterocyclic dyes based on pyrrole and
  thiophene: synthesis and evaluation of their optical, redox and photovoltaic
  properties}.
\newblock \bibinfo{journal}{Coatings} \bibinfo{volume}{12},
  \bibinfo{pages}{34}.
\bibitem[{Fernandes et~al.(2017)Fernandes, Castro, Pereira, Mendes, Serpa,
  Pina, Justino, Burrows and Raposo}]{Fernandes-2017}
\bibinfo{author}{Fernandes, S.S.}, \bibinfo{author}{Castro, M.C.R.},
  \bibinfo{author}{Pereira, A.I.}, \bibinfo{author}{Mendes, A.},
  \bibinfo{author}{Serpa, C.}, \bibinfo{author}{Pina, J.},
  \bibinfo{author}{Justino, L.L.}, \bibinfo{author}{Burrows, H.D.},
  \bibinfo{author}{Raposo, M.M.M.}, \bibinfo{year}{2017}.
\newblock \bibinfo{title}{Optical and photovoltaic properties of thieno [3,
  2-b] thiophene-based push--pull organic dyes with different anchoring groups
  for dye-sensitized solar cells}.
\newblock \bibinfo{journal}{ACS Omega} \bibinfo{volume}{2},
  \bibinfo{pages}{9268--9279}.
\bibitem[{Garcia-Amor{\'o}s et~al.(2016)Garcia-Amor{\'o}s, Castro, Coelho,
  Raposo and Velasco}]{garcia2016fastest}
\bibinfo{author}{Garcia-Amor{\'o}s, J.}, \bibinfo{author}{Castro, M.C.R.},
  \bibinfo{author}{Coelho, P.}, \bibinfo{author}{Raposo, M.M.M.},
  \bibinfo{author}{Velasco, D.}, \bibinfo{year}{2016}.
\newblock \bibinfo{title}{Fastest non-ionic azo dyes and transfer of their
  thermal isomerisation kinetics into liquid-crystalline materials}.
\newblock \bibinfo{journal}{Chemical Communications} \bibinfo{volume}{52},
  \bibinfo{pages}{5132--5135}.
\bibitem[{Garcia-Amor{\'o}s et~al.(2023)Garcia-Amor{\'o}s, Castro, Raposo and
  Velasco}]{garcia2023photochromic}
\bibinfo{author}{Garcia-Amor{\'o}s, J.}, \bibinfo{author}{Castro, M.C.R.},
  \bibinfo{author}{Raposo, M.M.M.}, \bibinfo{author}{Velasco, D.},
  \bibinfo{year}{2023}.
\newblock \bibinfo{title}{Photochromic heteroarylethenes with fast thermal
  isomerization kinetics}.
\newblock \bibinfo{journal}{Dyes and Pigments} \bibinfo{volume}{210},
  \bibinfo{pages}{111000}.
\bibitem[{Gluck and Agmon(2009)}]{gluck2009classical}
\bibinfo{author}{Gluck, P.}, \bibinfo{author}{Agmon, D.}, \bibinfo{year}{2009}.
\newblock \bibinfo{title}{Classical And Relativistic Mechanics}.
\newblock \bibinfo{publisher}{World Scientific Publishing Company}.
\bibitem[{Godoy~Alarcon et~al.(2021)Godoy~Alarcon, Santos and
  Bueno}]{Alarcon-2021}
\bibinfo{author}{Godoy~Alarcon, E.V.}, \bibinfo{author}{Santos, A.},
  \bibinfo{author}{Bueno, P.R.}, \bibinfo{year}{2021}.
\newblock \bibinfo{title}{Perspective on quantum electrochemistry. a simple
  method for measuring the electron transfer rate constant}.
\newblock \bibinfo{journal}{Electrochimica Acta} \bibinfo{volume}{398},
  \bibinfo{pages}{139219}.
\newblock \DOIprefix\doi{10.1016/j.electacta.2021.139219}.
\bibitem[{Gon{\c{c}}alves et~al.(2022)Gon{\c{c}}alves, Belmonte-Reche, Pina,
  Costa~da Silva, Pinto, Gallo, Costa and Raposo}]{gonccalves2022bioimaging}
\bibinfo{author}{Gon{\c{c}}alves, R.C.}, \bibinfo{author}{Belmonte-Reche, E.},
  \bibinfo{author}{Pina, J.}, \bibinfo{author}{Costa~da Silva, M.},
  \bibinfo{author}{Pinto, S.C.}, \bibinfo{author}{Gallo, J.},
  \bibinfo{author}{Costa, S.P.}, \bibinfo{author}{Raposo, M.M.M.},
  \bibinfo{year}{2022}.
\newblock \bibinfo{title}{Bioimaging of lysosomes with a bodipy ph-dependent
  fluorescent probe}.
\newblock \bibinfo{journal}{Molecules} \bibinfo{volume}{27},
  \bibinfo{pages}{8065}.
\bibitem[{Greiner et~al.(2000)}]{greiner2000relativistic}
\bibinfo{author}{Greiner, W.}, et~al., \bibinfo{year}{2000}.
\newblock \bibinfo{title}{Relativistic quantum mechanics}.
  volume~\bibinfo{volume}{2}.
\newblock \bibinfo{publisher}{Springer}.
\bibitem[{Jiang et~al.(2010)Jiang, Deng, Baba, Huang and
  Advincula}]{jiang2010monolayer}
\bibinfo{author}{Jiang, G.}, \bibinfo{author}{Deng, S.}, \bibinfo{author}{Baba,
  A.}, \bibinfo{author}{Huang, C.}, \bibinfo{author}{Advincula, R.C.},
  \bibinfo{year}{2010}.
\newblock \bibinfo{title}{On the monolayer adsorption of thiol-terminated
  dendritic oligothiophenes onto gold surfaces}.
\newblock \bibinfo{journal}{Macromolecular Chemistry and Physics}
  \bibinfo{volume}{211}, \bibinfo{pages}{2562--2572}.
\bibitem[{Keener et~al.(1968)Keener, Skelton and Snyder}]{keener1968synthesis}
\bibinfo{author}{Keener, R.L.}, \bibinfo{author}{Skelton, F.S.},
  \bibinfo{author}{Snyder, H.R.}, \bibinfo{year}{1968}.
\newblock \bibinfo{title}{Synthesis of 6-substituted thieno [3, 2-b] pyrroles}.
\newblock \bibinfo{journal}{The Journal of Organic Chemistry}
  \bibinfo{volume}{33}, \bibinfo{pages}{1355--1359}.
\bibitem[{Land and Horwitz(2022)}]{land2022relativistic}
\bibinfo{author}{Land, M.}, \bibinfo{author}{Horwitz, L.P.},
  \bibinfo{year}{2022}.
\newblock \bibinfo{title}{Relativistic classical mechanics and
  electrodynamics}.
\newblock \bibinfo{publisher}{Springer Nature}.
\bibitem[{Lopes et~al.(2024a)Lopes, Pinz{\'o}n, Dias-da Silva, Feliciano and
  Bueno}]{Lopes-2023}
\bibinfo{author}{Lopes, L.C.}, \bibinfo{author}{Pinz{\'o}n, E.},
  \bibinfo{author}{Dias-da Silva, G.}, \bibinfo{author}{Feliciano, G.T.},
  \bibinfo{author}{Bueno, P.R.}, \bibinfo{year}{2024}a.
\newblock \bibinfo{title}{Electrochemical measurement of the electronic
  structure of graphene via quantum mechanical rate spectroscopy}.
\newblock \bibinfo{journal}{Electrochimica Acta} , \bibinfo{pages}{143837}.
\bibitem[{Lopes et~al.(2024b)Lopes, Pinz{\'o}n, Dias-da Silva, Feliciano and
  Bueno}]{lopes2024electrochemical}
\bibinfo{author}{Lopes, L.C.}, \bibinfo{author}{Pinz{\'o}n, E.F.},
  \bibinfo{author}{Dias-da Silva, G.}, \bibinfo{author}{Feliciano, G.T.},
  \bibinfo{author}{Bueno, P.R.}, \bibinfo{year}{2024}b.
\newblock \bibinfo{title}{Electrochemical measurement of the electronic
  structure of graphene via quantum mechanical rate spectroscopy}.
\newblock \bibinfo{journal}{Electrochimica Acta} \bibinfo{volume}{480},
  \bibinfo{pages}{143837}.
\bibitem[{M.~Fernandes et~al.(2018)M.~Fernandes, Belsley, Pereira, Ivanou,
  Mendes, Justino, Burrows and Raposo}]{Fernandes-2018}
\bibinfo{author}{M.~Fernandes, S.S.}, \bibinfo{author}{Belsley, M.},
  \bibinfo{author}{Pereira, A.I.}, \bibinfo{author}{Ivanou, D.},
  \bibinfo{author}{Mendes, A.}, \bibinfo{author}{Justino, L.L.},
  \bibinfo{author}{Burrows, H.D.}, \bibinfo{author}{Raposo, M.M.M.},
  \bibinfo{year}{2018}.
\newblock \bibinfo{title}{Push--pull n, n-diphenylhydrazones bearing
  bithiophene or thienothiophene spacers as nonlinear optical second harmonic
  generators and as photosensitizers for nanocrystalline tio2 dye-sensitized
  solar cells}.
\newblock \bibinfo{journal}{ACS Omega} \bibinfo{volume}{3},
  \bibinfo{pages}{12893--12904}.
\bibitem[{Ma et~al.(2010)Ma, Yip, Huang and Jen}]{Ma-2010}
\bibinfo{author}{Ma, H.}, \bibinfo{author}{Yip, H.L.}, \bibinfo{author}{Huang,
  F.}, \bibinfo{author}{Jen, A.K.Y.}, \bibinfo{year}{2010}.
\newblock \bibinfo{title}{Interface engineering for organic electronics}.
\newblock \bibinfo{journal}{Advanced Functional Materials}
  \bibinfo{volume}{20}, \bibinfo{pages}{1371--1388}.
\newblock \DOIprefix\doi{10.1002/adfm.200902236}.
\bibitem[{Marcus(1964)}]{Marcus-1964}
\bibinfo{author}{Marcus, R.A.}, \bibinfo{year}{1964}.
\newblock \bibinfo{title}{Chemical and electrochemical electron-transfer
  theory}.
\newblock \bibinfo{journal}{Annu. Rev. Phys. Chem.} \bibinfo{volume}{15},
  \bibinfo{pages}{155}.
\newblock \DOIprefix\doi{10.1146/annurev.pc.15.100164.001103}.
\bibitem[{Marin-Hernandez et~al.(2014)Marin-Hernandez, Santos-Figueroa,
  Moragues, Raposo, Batista, Costa, Pardo, Martinez-Manez and
  Sancenon}]{marin2014imidazoanthraquinone}
\bibinfo{author}{Marin-Hernandez, C.}, \bibinfo{author}{Santos-Figueroa, L.E.},
  \bibinfo{author}{Moragues, M.E.}, \bibinfo{author}{Raposo, M.M.M.},
  \bibinfo{author}{Batista, R.M.}, \bibinfo{author}{Costa, S.P.},
  \bibinfo{author}{Pardo, T.}, \bibinfo{author}{Martinez-Manez, R.},
  \bibinfo{author}{Sancenon, F.}, \bibinfo{year}{2014}.
\newblock \bibinfo{title}{Imidazoanthraquinone derivatives for the
  chromofluorogenic sensing of basic anions and trivalent metal cations}.
\newblock \bibinfo{journal}{The Journal of Organic Chemistry}
  \bibinfo{volume}{79}, \bibinfo{pages}{10752--10761}.
\bibitem[{Meyer et~al.(2012)Meyer, Hamwi, Kroeger, Kowalsky, Riedl and
  Kahn}]{Mayer-2012}
\bibinfo{author}{Meyer, J.}, \bibinfo{author}{Hamwi, S.},
  \bibinfo{author}{Kroeger, M.}, \bibinfo{author}{Kowalsky, W.},
  \bibinfo{author}{Riedl, T.}, \bibinfo{author}{Kahn, A.},
  \bibinfo{year}{2012}.
\newblock \bibinfo{title}{Transition metal oxides for organic electronics:
  Energetics, device physics and applications}.
\newblock \bibinfo{journal}{Advanced Materials} \bibinfo{volume}{24},
  \bibinfo{pages}{5408--5427}.
\newblock \DOIprefix\doi{10.1002/adma.201201630}.
\bibitem[{Miranda and Bueno(2016)}]{Miranda-2016}
\bibinfo{author}{Miranda, D.A.}, \bibinfo{author}{Bueno, P.R.},
  \bibinfo{year}{2016}.
\newblock \bibinfo{title}{Density functional theory and an
  experimentally-designed energy functional of electron density}.
\newblock \bibinfo{journal}{Physical Chemistry Chemical Physics}
  \bibinfo{volume}{18}, \bibinfo{pages}{25984--25992}.
\newblock \DOIprefix\doi{10.1039/c6cp01659f}.
\bibitem[{Miranda and Bueno(2019)}]{Miranda-2019}
\bibinfo{author}{Miranda, D.A.}, \bibinfo{author}{Bueno, P.R.},
  \bibinfo{year}{2019}.
\newblock \bibinfo{title}{Chemical hardness of mesoscopic electrochemical
  systems directly analyzed from experimental data}.
\newblock \bibinfo{journal}{Journal of Physical Chemistry C}
  \bibinfo{volume}{123}, \bibinfo{pages}{21213--21223}.
\newblock \DOIprefix\doi{10.1021/acs.jpcc.9b04020}.
\bibitem[{Novoselov et~al.(2005)Novoselov, Geim, Morozov, Jiang, Katsnelson,
  Grigorieva, Dubonos and Firsov}]{Novoselov2005a}
\bibinfo{author}{Novoselov, K.S.}, \bibinfo{author}{Geim, A.K.},
  \bibinfo{author}{Morozov, S.V.}, \bibinfo{author}{Jiang, D.},
  \bibinfo{author}{Katsnelson, M.I.}, \bibinfo{author}{Grigorieva, I.V.},
  \bibinfo{author}{Dubonos, S.V.}, \bibinfo{author}{Firsov, A.A.},
  \bibinfo{year}{2005}.
\newblock \bibinfo{title}{Two-dimensional gas of massless dirac fermions in
  graphene}.
\newblock \bibinfo{journal}{Nature} \bibinfo{volume}{438},
  \bibinfo{pages}{197--200}.
\newblock \URLprefix \url{https://doi.org/10.1038/nature04233},
  \DOIprefix\doi{10.1038/nature04233}.
\bibitem[{Oliva et~al.(2006)Oliva, Casado, Raposo, Fonseca, Hartmann,
  Hern{\'a}ndez and L{\'o}pez~Navarrete}]{oliva2006structure}
\bibinfo{author}{Oliva, M.M.}, \bibinfo{author}{Casado, J.},
  \bibinfo{author}{Raposo, M.M.M.}, \bibinfo{author}{Fonseca, A.M.C.},
  \bibinfo{author}{Hartmann, H.}, \bibinfo{author}{Hern{\'a}ndez, V.},
  \bibinfo{author}{L{\'o}pez~Navarrete, J.T.}, \bibinfo{year}{2006}.
\newblock \bibinfo{title}{Structure- property relationships in push- pull
  amino/cyanovinyl end-capped oligothiophenes: Quantum chemical and
  experimental studies}.
\newblock \bibinfo{journal}{The Journal of Organic Chemistry}
  \bibinfo{volume}{71}, \bibinfo{pages}{7509--7520}.
\bibitem[{Piz{\'o}n et~al.(2024)Piz{\'o}n, Lopes, Fonseca, Schiavon and
  Bueno}]{pinzon2023quantum}
\bibinfo{author}{Piz{\'o}n, E.F.}, \bibinfo{author}{Lopes, L.C.},
  \bibinfo{author}{Fonseca, A.F.V.}, \bibinfo{author}{Schiavon, M.A.},
  \bibinfo{author}{Bueno, P.R.}, \bibinfo{year}{2024}.
\newblock \bibinfo{title}{Quantum rate as a spectroscopic methodology for
  measuring the electronic structure of quantum dots}.
\newblock \bibinfo{journal}{Journal of Materials Chemistry C} .
\bibitem[{Raposo et~al.(2006)Raposo, Sousa, Kirsch, Cardoso, Belsley,
  de~Matos~Gomes and Fonseca}]{raposo2006synthesis}
\bibinfo{author}{Raposo, M.M.M.}, \bibinfo{author}{Sousa, A.M.},
  \bibinfo{author}{Kirsch, G.}, \bibinfo{author}{Cardoso, P.},
  \bibinfo{author}{Belsley, M.}, \bibinfo{author}{de~Matos~Gomes, E.},
  \bibinfo{author}{Fonseca, A.M.C.}, \bibinfo{year}{2006}.
\newblock \bibinfo{title}{Synthesis and characterization of
  dicyanovinyl-substituted thienylpyrroles as new nonlinear optical
  chromophores}.
\newblock \bibinfo{journal}{Organic Letters} \bibinfo{volume}{8},
  \bibinfo{pages}{3681--3684}.
\bibitem[{Sanchez et~al.(2022)Sanchez, Santos and Bueno}]{Sanchez-2022-1}
\bibinfo{author}{Sanchez, Y.P.}, \bibinfo{author}{Santos, A.},
  \bibinfo{author}{Bueno, P.R.}, \bibinfo{year}{2022}.
\newblock \bibinfo{title}{Quantum mechanical meaning of the charge transfer
  resistance}.
\newblock \bibinfo{journal}{Journal of Physical Chemistry C}
  \bibinfo{volume}{126}, \bibinfo{pages}{3151--3162}.
\newblock \DOIprefix\doi{10.1021/acs.jpcc.1c07801}.
\bibitem[{S{\'a}nchez et~al.(2022)S{\'a}nchez, Santos and
  Bueno}]{sanchez2022quantum}
\bibinfo{author}{S{\'a}nchez, Y.P.}, \bibinfo{author}{Santos, A.},
  \bibinfo{author}{Bueno, P.R.}, \bibinfo{year}{2022}.
\newblock \bibinfo{title}{Quantum mechanical meaning of the charge transfer
  resistance}.
\newblock \bibinfo{journal}{The Journal of Physical Chemistry C}
  \bibinfo{volume}{126}, \bibinfo{pages}{3151--3162}.
\bibitem[{Santos et~al.(2020)Santos, Tefashe, McCreery and
  Bueno}]{santos2020introducing}
\bibinfo{author}{Santos, A.}, \bibinfo{author}{Tefashe, U.M.},
  \bibinfo{author}{McCreery, R.L.}, \bibinfo{author}{Bueno, P.R.},
  \bibinfo{year}{2020}.
\newblock \bibinfo{title}{Introducing mesoscopic charge transfer rates into
  molecular electronics}.
\newblock \bibinfo{journal}{Physical Chemistry Chemical Physics}
  \bibinfo{volume}{22}, \bibinfo{pages}{10828--10832}.
\bibitem[{Tobias(2023)}]{Tobias-2023}
\bibinfo{author}{Tobias, B.}, \bibinfo{year}{2023}.
\newblock \bibinfo{title}{First-principles theory of electrochemical
  capacitance}.
\newblock \bibinfo{journal}{Electrochimica Acta} \bibinfo{volume}{444},
  \bibinfo{pages}{142016}.
\newblock \DOIprefix\doi{10.1016/j.electacta.2023.142016}.
\bibitem[{Welch and Phillips(1999)}]{welch1999improved}
\bibinfo{author}{Welch, M.}, \bibinfo{author}{Phillips, R.S.},
  \bibinfo{year}{1999}.
\newblock \bibinfo{title}{Improved syntheses of [3, 2-b]-and [2, 3-b]-fused
  selenolo-and thienopyrroles, and of furo [3, 2-b] pyrrole}.
\newblock \bibinfo{journal}{Heterocyclic Communications} \bibinfo{volume}{5},
  \bibinfo{pages}{305--310}.

\end{thebibliography}

\end{document}